\documentclass[draftclsnofoot,onecolumn,12pt]{IEEEtran}

\usepackage{graphicx}
\usepackage{epstopdf}

\usepackage{subfigure}
\usepackage{booktabs}
\usepackage{caption}
\usepackage{color}
\usepackage{bm}
\usepackage{CJK}
\usepackage{indentfirst}
\usepackage{amsmath}
\usepackage{amssymb}
\usepackage{float}
\usepackage{multirow}
\usepackage{algorithm}
\usepackage{algpseudocode}
\usepackage{amsmath}
\usepackage{flushend}
\usepackage{geometry}
\newtheorem{lemma}{Lemma}
\newtheorem{proposition}{Proposition}

\begin{document}
\title{Joint Spatial Division and Multiplexing with Customized Orthogonal Group Channels in Multi-RIS-Assisted Systems}
\author{\IEEEauthorblockN{Weicong~Chen, Chao-Kai~Wen, Wankai~Tang, Xiao~Li, and~Shi~Jin}\\
	\thanks{{Weicong Chen, Wankai Tang, Xiao Li, and Shi Jin are with the National Mobile Communications Research Laboratory, Southeast University, Nanjing, 210096, P. R. China (e-mail: cwc@seu.edu.cn; tangwk@seu.edu.cn; li\_xiao@seu.edu.cn; jinshi@seu.edu.cn).} }
\thanks{Chao-Kai Wen is with the Institute of Communications Engineering, National Sun Yat-sen University, Kaohsiung 80424, Taiwan. (e-mail: chaokai.wen@mail.nsysu.edu.tw).}
}

\maketitle
\begin{abstract}
Reconfigurable intelligent surfaces (RISs) offer the unique capability to reshape the radio environment, thereby simplifying transmission schemes traditionally contingent on channel conditions. Joint spatial division and multiplexing (JSDM) emerges as a low-overhead transmission scheme for multi-user equipment (UE) scenarios, typically requiring complex matrix decomposition to achieve block-diagonalization of the effective channel matrix. In this study, we introduce an innovative JSDM design that leverages RISs to customize channels, thereby streamlining the overall procedures. By strategically positioning RISs at the discrete Fourier transform (DFT) directions of the base station (BS), we establish orthogonal line-of-sight links within the BS-RIS channel, enabling a straightforward pre-beamforming design. \textcolor{black}{Based on UE grouping, we devise reflected beams of the RIS with optimized directions to mitigate inter-group interference in the RISs-UEs channel.} An approximation of the channel cross-correlation coefficient is derived and serves as a foundation for the RISs-UEs association, further diminishing inter-group interference. Numerical results substantiate the efficacy of our RIS-customized JSDM in not only achieving effective channel block-diagonalization but also in significantly enhancing the sum spectral efficiency for multi-UE transmissions.
\end{abstract}

\section{Introduction}
\IEEEPARstart{D}{riven} by usage scenarios such as massive communications, ubiquitous connectivity, and integrated sensing and communication, the landscape of mobile systems is undergoing a transformative phase \cite{imt-2030}. Central to the efficient design of wireless transmission schemes in multi-user massive multiple-input-multiple-output \textcolor{black}{(MIMO)} systems is the acquisition of channel state information \textcolor{black}{(CSI)} \cite{CE-ALS} \cite{CE-CC}. In time division duplex \textcolor{black}{(TDD)} systems, the reciprocity of uplink and downlink channels allows the base station (BS) to acquire downlink CSI from uplink transmissions \cite{TDD-1}, \cite{TDD-2}. However, in frequency division duplex \textcolor{black}{(FDD)} systems, the absence of channel reciprocity necessitates the BS to obtain downlink CSI via uplink feedback from the user equipment \textcolor{black}{(UE)} \cite{FDD-FD-1}--\cite{FDD-FD-4}. \textcolor{black}{When the number of BS antennas significantly increases, feedback of the high-dimensional channel incurs substantial overhead, diminishing the efficiency of wireless transmission. In this context, joint spatial division and multiplexing (JSDM) \cite{JSDM}, which utilizes the pre-beamforming (PBF) \footnote{\textcolor{black}{The PBF is a technique used to pre-process signals before transmission. In JSDM, it is designed with the statistical CSI (S-CSI) and aims to reduce inter-group interference by aligning the signal space with specific user groups.}} to reduce the dimension of the multi-UE channel, emerges as an attractive solution.} 

The implementation of JSDM confronts several challenges, primarily stemming from the reliance on the spatial angular distribution of scattering environments in UE groups relative to the BS. The uncontrollable nature of wireless channels places JSDM design at the mercy of the ever-changing propagation environment, \textcolor{black}{which necessitates that the BS has a large-scale antenna array to achieve orthogonality between channels of UE groups \cite{JSDM}. Moreover, the adoption of a fully-digital antenna array, \textcolor{black}{where each antenna element is individually connected to a digital baseband processing unit,} incurs considerable hardware costs and power consumption \cite{fully-digital}, which is an undesirable prospect for future systems that demand low-cost and efficient communication.} \textcolor{black}{The introduction of a reconfigurable intelligent surface (RIS) presents a timely and effective counter to the challenges of JSDM. Specifically, the RIS constitutes a segment of the channel between the transceiver, thereby enabling channel controllability. Typically consisting of a large array of reflecting elements, the RIS can achieve orthogonality of sub-channels, similar to that achieved by large-scale BS antennas. Benefiting from the absence of RF chains, the RIS can significantly reduce hardware costs and power consumption.}

\textcolor{black}{The RIS application has been effectively studied in various wireless communication scenarios. In multi-UE contexts, research on single-RIS \cite{24-Deploy}--\cite{JSDM-1RIS} and multi-RIS \cite{JSDM-ICC}--\cite{21-I-2RIS} assistance has been extensive. Taking into account the fixed position of an assistant RIS and the variations in locations of UEs, an RIS deployment strategy based on the long-term geographic distribution of UEs was proposed in \cite{24-Deploy} to maximize the sun rate of UEs.	In a vehicular scenario, the authors in \cite{22-GC-deploy} identified an optimal location, height, and downtile for the RIS to counteract blockage-induced link vulnerabilities by accounting for the distribution of potential UEs and possible blockers. Hardware impairments and phase noise in RIS are considered for maximizing multi-UE rates in \cite{22-S-CPan}. For the first time, the JSDM was employed in RIS-assisted FDD systems \cite{JSDM-1RIS}, using deterministic equivalent analysis to derive sum spectral efficiency (SE).} However, single-RIS systems often fail to support multi-UE services in far-field line-of-sight (LoS) scenarios due to the rank-1 nature of the BS--RIS channel, limiting multi-stream transmission despite the presence of a massive antenna array.

\textcolor{black}{The introduction of multi-RIS enhances spatial multiplexing for multi-UE transmissions by providing a greater degree of freedom in system design. To overcome the rank deficiency of a single-RIS-assisted JSDM, the authors in \cite{JSDM-ICC} proposed to install distributed RISs between the BS and UEs, increasing the rank of the channel between the BS and RISs. The two-timescale beamforming and multi-RIS-assisted JSDM were combined in \cite{JSDM-TWC}, where UEs in each group were served via randomized beamforming in different time resources.	The application of multi-RIS in unmanned aerial vehicle-based multi-UE downlink communications was explored in \cite{23-S-AB}, which focused on enhancing network coverage and reducing bit error rates via the selection of RISs. To guarantee unobstructed LoS channels between transceivers in indoor THz communication systems, the authors in \cite{THz-deploy} optimized the placement of RISs to maximize the coverage area. Research by  authors in \cite{21-I-2RIS} demonstrated that a double-RIS assisted multi-UE system} surpasses single-RIS systems in performance by increasing the effective channel rank. \textcolor{black}{The aforementioned studies focused on the optimization of RIS selection \cite{23-S-AB}, deployment \cite{THz-deploy},  beam scheduling \cite{JSDM-TWC}, and reflection coefficients \cite{JSDM-ICC}, \cite{21-I-2RIS} under various performance objectives, but did not reveal how multiple RIS can enhance the transmission performance of multi-UE systems by reshaping the wireless channel.}

The performance gains of the RIS are achieved primarily by the optimization of its reflection coefficients. The design methodologies for the RIS and the consequent performance gains vary depending on whether the instantaneous CSI \textcolor{black}{(I-CSI)} or the S-CSI is utilized \cite{app-6}. When the I-CSI is employed, the RIS design can be highly responsive to the varying propagation environment. In this case, the signal quality can be optimized in near real-time, potentially leading to significant improvement in performance such as sum SE \cite{opt-1-SE}, energy efficiency \cite{opt-2-EE}, and outage probability \cite{opt-3-op}. However, obtaining the I-CSI is extremely complex and resource-intensive in RIS-assisted systems, stemming from the low-cost requirement that strips signal processing capability from the RIS. In cases where only the S-CSI is available, the RIS design relies on long-term channel characteristics, such as the  direction of signal paths \cite{FDD-FD-3} and the spatial correlation of the channel \cite{JSDM-1RIS,opt-4}. While this approach may not achieve the same performance gain as those with I-CSI, it is more practical due to the simplification of the RIS design and the lower requirement for CSI acquisition.

This study proposes a low-overhead JSDM approach for multi-UE transmission in RIS-assisted MIMO systems, employing S-CSI-based PBF to reduce the dimension of the required I-CSI. \textcolor{black}{The first application of RIS to JSDM was introduced in \cite{JSDM-1RIS}, utilizing a single RIS to assist the BS. However, in typical RIS deployment scenarios where the RIS is installed within the far-field LoS range of the BS, the BS-RIS channel becomes severely rank-deficient, rendering it incapable of supporting multi-stream transmission. To address this issue, the authors in \cite{JSDM-ICC} proposed a distributed RIS architecture. In their two-hop RIS reflection model, the signal from the BS is first reflected by multiple intermediate RISs before being directed to all UEs by a main RIS. Although this scheme effectively resolves the rank deficiency of the BS-RIS channel, it introduces more severe path loss. Moreover, when multiple UEs are in close proximity, the RIS-UE channel still suffers from rank deficiency. The authors in \cite{JSDM-TWC} further employed different RISs to serve different UE groups, reducing the correlation between UEs in different groups and achieving fair transmission for UEs within the same group on different time resources through randomized RIS beamforming. Unlike the conventional JSDM approaches in \cite{JSDM-1RIS} and \cite{JSDM-ICC}, which rely on matrix decomposition to design PBF for block-diagonalizing the effective channel,  we propose a low-complexity method that configures RISs with the direction of channel paths to customize orthogonal UE group channels. To serve UEs within the same group simultaneously, rather than on different time resources as in \cite{JSDM-TWC}, we introduce multiple RISs for each group of UEs. Furthermore, we address the challenges of grouping UEs and RISs, as well as the RIS-UE association problem, which were not covered in \cite{JSDM-1RIS}--\cite{JSDM-TWC}.} The contributions of our work are summarized as follows: 

\begin{itemize}
\item {A low-complexity block diagonalization (BD) for the effective multi-UE channel.} \textcolor{black}{Differing from \cite{JSDM-1RIS} and \cite{JSDM-ICC}, which follow the classic analysis procedure of JSDM, we streamline the BD of the effective channel by innovatively utilizing the channel customization capability of RISs.} Specifically, in far-field scenarios, we deploy RISs at the discrete Fourier transform (DFT) directions of the BS antenna array to create orthogonal LoS links from the BS to each RIS while significantly reducing the design complexity of the PBF. \textcolor{black}{By steering the reflected beams of RISs toward intended UE groups with optimized directions, the inter-group interferences are reduced, resulting in an approximated block-diagonal effective multi-UE channel.}

\item \textcolor{black}{{Reducing implementation complexity and cost of the PBF for the JSDM.} \textcolor{black}{The proposed PBF is determined by the deployment locations of RISs and does not require updates when RISs are fixed.} In contrast to conventional JSDM schemes, its design process does not involve high-dimensional matrix decomposition. The PBF matrix, constructed from DFT vectors, can be effectively implemented on a hybrid antenna architecture with a number of RF chains equivalent to the number of UEs, while achieving performance comparable to that of a fully-digital beamforming architecture. Compared to conventional JSDM schemes that require large-scale digital antenna architectures, this approach significantly reduces hardware costs and power consumption.} 

\item {Grouping and association of UEs and RISs for reducing channel correlation.} On the basis of the proposed JSDM with the customized channel, we derive an approximation of the channel cross-correlation coefficient for different UEs. This approximation shows that the channel cross-correlation coefficient is determined by the direction and distance of each LoS path in the RISs--UEs channel, providing a fundamental principle when dividing UEs and RISs into groups to further reduce inter-group interference. When the number of UEs exceeds the channel rank, a UE-RIS association algorithm is proposed to assign RISs to the effective UEs, with the aim of channel correlation minimization.

\item {Comprehensive simulations combined with insightful discussion about the proposed JSDM.} Numerical simulations are conducted to show the effectiveness of the proposed low-complexity BD for the effective channel. With optimized beam directions, the proposed JSDM with the customized channel can adaptively achieve better sum SE in the noise- and interference-limited regime, respectively. By strategically selecting the number of activated RISs, a balance between serving a greater number of UEs and achieving a higher sum SE can be struck. We show that our proposal is robust against the Rician factor.

\end{itemize}

The remainder of this paper is structured as follows. Section \ref{sec:2} provides the foundational system model for this research. Section \ref{sec:3} introduces the novel JSDM with RIS-customized channel. The channel correlation is analyzed in Section \ref{sec:4} for UE-RIS association. Section \ref{sec:5} presents a comprehensive analysis of the numerical results validating our proposal. Finally, Section \ref{sec:6} concludes this study.

\emph{Notations}: Lowercase and uppercase of bold letters represent vectors and matrices, respectively. Superscripts $(\cdot)^T$ and $(\cdot)^H$ denote transpose and conjugate transpose operations. ${\rm tr}(\cdot)$ calculates the trace of a matrix. ${\mathbb C}^{a\times b}$ represents the set of $a\times b$ complex matrices. $|\cdot|$ and $\|\cdot\|$ indicate the absolute value and Euclidean norm. The Kronecker product is denoted by $\otimes$. ${\mathbb E}\{\cdot\}$ calculates statistical expectation. ${\rm blkdiag}\{{\bf X}_1,{\bf X}_2,\dots,{\bf X}_N\}$ represents a block diagonal matrix with matrices $\{ {\bf X}_i\}$. ${\rm diag}(a_1,a_2,\dots,a_N)$ forms a diagonal matrix with diagonal elements $\{a_i\}$. Table \ref{Tab:1} provides the definition of main parameters.
 
\begin{table}[h]
	\caption{Definition of main parameters}
	\centering
	\label{Tab:1}
	\begin{tabular}{ll}
		\toprule
		Parameter & Description\\
		\midrule    
		${\mathcal N}_c$/${\mathcal K}_c$ & set of indexes of UEs/RISs in group $c$\\
		${\bf h}^{\rm BU}_{n}$& the BS--UE$_n$ channel matrix\\
		${\bf B}_k$ & the BS-RIS$_k$ channel matrix\\
		${\bf u}_{k,n}$ & the RIS$_k$--UE$_n$ channel\\
		${\bf U}_{k}$ & the RIS$_k$--UEs channel\\
		${\bf r}_{k,n,l}$& ARV at the RIS for path $l$ in the RIS$_k$--UE$_n$ channel\\
		${\bf b}_{k,l}$/${\bf r}_{k,l}$& ARV at the BS/RIS for path $l$ in the BS--RIS$_k$ channel\\
		$\theta _{k,l}^{{\rm{B,D}}}$, $\phi _{k,l}^{{\rm{B,D}}}$ & AoDs of path $l$ in the BS--RIS$_k$ channel\\
		$\theta _{k,l}^{{\rm{B,A}}}$, $\phi _{k,l}^{{\rm{B,A}}}$ & AoAs of path $l$ in the BS--RIS$_k$ channel\\
		$\theta _{k,n,l}^{{\rm{U,D}}}$, $\phi _{k,n,l}^{{\rm{U,D}}}$ & AoDs of path $l$ in the RIS$_k$--UE$_n$ channel\\
		$\theta _{k,n,l}^{{\rm{U,A}}}$, $\phi _{k,n,l}^{{\rm{U,A}}}$ & AoAs of path $l$ in the RIS$_k$--UE$_n$ channel\\
		 ${\theta _{k,c}^{\rm cntr}}$/${\phi _{k,c}^{\rm cntr}}$ & AoDs from RIS$_k$ to the center of the group $c$\\
		\bottomrule
	\end{tabular}
\end{table}


\section{System Model}\label{sec:2}

\begin{figure}
	\centering
	\includegraphics[width=0.5\textwidth]{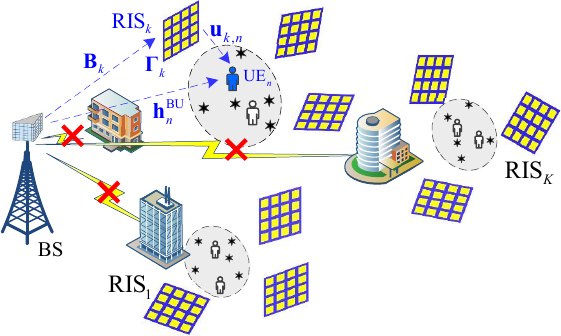}
	\caption{Multi-UE MIMO system assisted by multiple RISs.}
	\label{Fig.system-model} 
\end{figure}

We consider a multi-RIS-assisted multi-UE MIMO system operating in the FDD mode. As shown in Fig. \ref{Fig.system-model}, several coverage holes exist in the system due to blockages. $K$ RISs, denoted by ${\mathcal K}$, are mounted on the LoS between the BS and these coverage holes to establish reliable reflection/refraction links and restore communication services without constructing new BSs\footnote{\textcolor{black}{The control, coordination, and time synchronization of multiple RISs are crucial for achieving the full potential of RIS-assisted systems and have been studied in \cite{coord} and \cite{time-syn}. In the considered scenarios, where the BS and RISs are fixed at specific locations, RISs can be reliably controlled by the BS and synchronized with the BS via solutions utilizing the global navigation satellite system \cite{GPS}.}}. In this system, the BS and the RIS are equipped with a uniform planar array \textcolor{black}{(UPA)} with half-wavelength element spacing. Each UE has a single antenna. The BS\footnote{In the following, superscripts with uppercase text B, R, U, A, D, L, and N represent the BS, RIS, UE, angle of Arrival, angle of Departure, Line-of-sight, and Non-line-of-sight, respectively.} possesses $M^{\rm B}=M^{\rm B,v}\times M^{\rm B,h}$ transmit antennas, and each RIS comprises $M^{\rm R}=M^{\rm R,v}\times M^{\rm S,h}$ elements. Here, superscripts ${\rm v}$ and ${\rm h}$ denote the vertical and horizontal numbers of antennas/elements, respectively. We assume $N$ UEs, denoted by ${\mathcal N}$, are located within coverage holes. Considering that UEs within the same coverage hole share similar spatial distribution properties, we divided both UEs and RISs into $C$ groups. In group $c$, there are $N_c$ UEs serviced by $K_c$ RISs, where $N = \sum\nolimits_{c = 1}^C {{N_c}} $ and $K = \sum\nolimits_{c = 1}^C {{K_c}}$. The sets of UEs and RISs belonging to group $c$ are denoted by ${\mathcal N }_c\subset {\mathcal N}$ and ${\mathcal K }_c\subset {\mathcal K}$, respectively.

The downlink received signal is given by
\begin{equation}
	{\bf y} = {\bf H}^H {\bf x} + {\bf z},
\end{equation}
where ${\bf y}\in {\mathbb{C}}^{N}$ is the stacked signal of $N$ UEs; ${\bf{H}} = [ {{{\bf{H}}_1}, \dots ,{{\bf{H}}_C}} ]\in {\mathbb{C}}^{M^{\rm B}\times N}$ with the channel for UEs in group $c$ being ${\bf H}_c \in {\mathbb{C}}^{M^{\rm B}\times N_{c}}$, whose columns are $\{{\bf h}_n ,n\in {\mathcal N}_c\}$; ${\bf x}\in {\mathbb{C}}^{M^{\rm B}}$ is the transmit signal; and ${\bf z}\sim {\mathcal {CN}}(0,\sigma^2 {\bf I}_{N})$ is the additive white Gaussian noise with power $\sigma^2$, where ${\bf I}_{N}$ represents the identity matrix with the dimension being $N\times N$.

\subsection{Channel Model}

The end-to-end channel ${\bf h}_n$ for UE$_n$, comprising the blocked direct BS--UE channel and the cascaded BS--RISs--UE channel\footnote{\textcolor{black}{Owing to the multiplicative cascaded path loss, which has been theoretically established in \cite[Eq. (6)]{pathloss-1} and \cite[Eq. (18)]{pathloss-2} and experimentally confirmed in \cite{pathloss-1}, the signal components associated with inter-RIS reflection are considerably weaker than those reflected by a single RIS. Therefore, it is justifiable to consider only the channels cascaded by a single RIS and neglect those cascaded by multiple RISs.}}, can be expressed as \cite{channel-model}
\begin{equation}
	{{\bf{h}}_n} = {{\bf{h}}^{\rm BU}_{n}}+ \sum\limits_{k = 1}^K {{{\bf{B}}_k}{{{\bm{\Gamma }}_k}} {{\bf{u}}_{k,n}}},
\end{equation}
where ${{\bf{h}}^{\rm BU}_{n}}\sim {\mathcal{CN}}(0,{(\epsilon^{\rm BU}_{n}\rho^{\rm BU}_{n})^2 }{\bf I}_{M^{\rm B}})$  is the blocked direct channel from the BS to UE$_n$ with penetration loss $\epsilon^{\rm BU}_{n}$ and path loss $\rho^{\rm BU}_{n}= {\lambda}/{(4\pi d^{\rm BU}_{n})}$, where $d^{\rm BU}_{n}$ is the distance from the BS to UE$_n$; ${{\bf \Gamma}_k}={\rm diag}({\bm \gamma}_k)$ with ${\bm \gamma_k}=[\gamma_{k,1},\dots,\gamma_{k,M^{\rm R}}]^T \textcolor{black}{\in {\mathbb{C}}^{M^{\rm R}\times1}}$ is the reflection coefficient matrix of RIS$_k$, ${\bf B}_k\textcolor{black}{\in {\mathbb{C}}^{M^{\rm B}\times{M^{\rm R}}}}$ and ${\bf u}_{k,n}\textcolor{black}{\in {\mathbb{C}}^{M^{\rm R}\times1}}$ are the BS--RIS$_k$ and RIS$_k$--UE$_n$ channel, respectively.

To operate efficiently, it is advantageous for RISs to establish LoS links both toward the BS and the UEs \cite{Rician}. Thus, we adopt the Rician channel model for ${\bf B}_k$ and ${\bf u}_{k,n}$. The BS--RIS$_k$ channel is modeled as
\begin{equation}
	{\bf B}_k = {\rho_{k}^{\rm B}}\left(\sqrt{\frac{\kappa^{\rm B}}{\kappa^{\rm B}+1}}{\bf B}^{{\rm L}}_k + \sqrt{\frac{1}{\kappa^{\rm B}+1}}{\bf B}^{{\rm N}}_k\right),
\end{equation}
where $ {\rho_{k}^{\rm B}}= {\lambda}/{(4\pi d_k^{\rm B})}$ is the path loss with $d_{k}^{\rm B}$ being the distance from the BS to RIS$_k$, $\kappa^{\rm B}$ is the Rician factor, ${\bf B}^{{\rm L}}_k$ and ${\bf B}^{{\rm N}}_k$ are the LoS and non-LoS (NLoS) channel components of ${\bf B}_k$, respectively. In the far-field, the LoS component can be expressed as
\begin{equation}
	\textcolor{black}{{{\bf{B}}^{{\rm{L}}}_k} = \sqrt {{M^{\rm{B}}}{M^{\rm R}}} \ e^{{j{\mu^{\rm B}_{k}}}} {\bf{b}}{\left( {\theta _{k,0}^{{\rm{B,D}}},\phi _{k,0}^{{\rm{B,D}}}} \right)} {{\bf{r}}^H}{\left( {\theta _{k,0}^{{\rm{B,A}}},\phi _{k,0}^{{\rm{B,A}}}} \right)}},
\end{equation}
where \textcolor{black}{$\mu^{\rm B}_{k}={2\pi d_{k}^{\rm B}}/{\lambda}$ is the common phase shift introduced by the LoS distance $d_{k}^{\rm B}$,} $\theta _{k,0}^{{\rm{B,D}}}$ (or $\theta _{k,0}^{{\rm{B,A}}}$) and $\phi _{k,0}^{{\rm{B,D}}}$ (or $\phi _{k,0}^{{\rm{B,A}}}$) denote respectively the vertical and horizontal angle-of-departure (AoD) (or angle-of-arrival (AoA)) of the LoS link,  ${\bf b}(\cdot) \textcolor{black}{\in {\mathbb{C}}^{M^{\rm B}\times1}}$ and ${\bf r}(\cdot) \textcolor{black}{\in {\mathbb{C}}^{M^{\rm R}\times1}}$ are the array response vector (ARV) at the BS and the RIS, respectively.  The ARV of a UPA can be represented by a unified form:
\begin{equation}\label{eq:ARV-a}
	\begin{aligned}
		{\bf{a}}\left( {\theta ,\phi } \right) = \frac{1}{{\sqrt M }}{\left[ {1,{e^{j\pi \cos \theta }}, \dots ,{e^{j\left( {{M^{\rm{v}}}{\rm{ - }}1} \right)\pi \cos \theta }}} \right]^T} \\
		\otimes {\left[ {1,{e^{j\pi \sin \theta \cos \phi }}, \dots ,{e^{j\left( {{M^{\rm{h}}}{\rm{ - }}1} \right)\pi \sin \theta \cos \phi }}} \right]^T},
	\end{aligned}
\end{equation}
where $M$ is the total element number with $M^{\rm v}$ and $M^{\rm h}$ being the vertical and horizontal part, respectively. The geometry multipath channel model \cite{NLoS_model} is adopted for ${\bf B}^{{\rm N}}_k$, which can be expressed as
\begin{equation}
		{{\bf{B}}^{{\rm{N}}}_k} = \sqrt {\frac{{{M^{\rm{B}}}{M^{\rm R}}}}{{{L_k}}}} \sum\limits_{l = 1}^{{L_k}} {\eta _{k,l}^{\rm{B}}{\bf{b}}{\left( {\theta _{k,l}^{{\rm{B,D}}},\phi _{k,l}^{{\rm{B,D}}}} \right)}{{\bf{r}}^H}{\left( {\theta _{k,l}^{{\rm{B,A}}},\phi _{k,l}^{{\rm{B,A}}}} \right)}},
\end{equation}
where $L_{k}$ is the number of NLoS paths, $\eta_{k,l}^{\rm B}\sim {\mathcal{CN}}(0,1)$ is the complex path gain of the $l$-th NLoS path\footnote{\textcolor{black}{Given that the LoS channel has a much longer coherence time than the NLoS channel, the parameters of the LoS and NLoS channels in this study are defined as S-CSI and I-CSI, respectively. Since S-CSI, which includes the direction and distance of the LoS path, can be calculated from the positions of the BS, RISs, and UEs, it is reciprocal between the uplink and downlink. This reciprocity enables the BS to easily obtain S-CSI through localization techniques \cite{F-RIS}.}}. By denoting index of the LoS path as $0$, \textcolor{black}{$\beta _{k,0}^{\rm{B}} = \sqrt {\frac{{{M^{\rm{B}}}{M^{\rm R}}{\kappa ^{\rm{B}}}}}{{{\kappa ^{\rm{B}}} + 1}}} \rho _k^{\rm{B}}e^{j\mu^{\rm B}_{k}}$,} and $\beta _{k,l}^{\rm{B}} = \sqrt {\frac{{{M^{\rm{B}}}{M^{\rm R}}}}{{( {{\kappa ^{\rm{B}}} + 1} ){L_k}}}} \rho _k^{\rm{B}}\eta _{k,l}^{\rm{B}}$ ($l>0$), the BS--RIS$_k$ channel can be unified as
\begin{equation}
	{{\bf{B}}_k}=\sum\limits_{l = 0}^{{L_k}} {\beta _{k,l}^{\rm{B}}{\bf{b}}_{k,l}{{\bf{r}}^H_{k,l}}} ,
\end{equation}
where ${\bf b}_{k,l}={\bf{b}}( {\theta _{k,l}^{{\rm{B,D}}},\phi _{k,l}^{{\rm{B,D}}}} )$ and ${\bf r}_{k,l}={{\bf{r}}}( {\theta _{k,l}^{{\rm{B,A}}},\phi _{k,l}^{{\rm{B,A}}}} )$.

In analogy with ${{\bf{B}}_k}$, RIS$_k$--UE$_n$ channel is modeled as
\begin{equation}
		{{\bf{u}}_{k,n}}=\sum\limits_{l = 0}^{{L_{k,n}}} {\beta _{k,n,l}^{\rm{U}}{\bf{r}}_{k,n,l}},
\end{equation}
where ${\bf r}_{k,n,l}$ denotes ${\bf{r}}( {\theta _{k,n,l}^{{\rm{U,D}}},\phi _{k,n,l}^{{\rm{U,D}}}} )$, $L_{k,n}$ is the number of NLoS paths; \textcolor{black}{$\beta _{k,n,0}^{\rm{U}} = \sqrt {\frac{{{M^{\rm R}}{\kappa ^{\rm{U}}}}}{{{\kappa ^{\rm{U}}} + 1}}} \rho _{k,n}^{\rm{U}}e^{j\mu^{\rm U}_{k,n}}$ and $\beta _{k,n,l}^{\rm{U}} = \sqrt {\frac{{{M^{\rm R}}}}{{( {{\kappa ^{\rm{U}}} + 1} ){L_{k,n}}}}} \rho _{k,n}^{\rm{U}}\eta _{k,n,l}^{\rm{U}}$ ($l>0$),} $\rho _{k,n}^{\rm{U}}= {\lambda}/{(4\pi d_{k,n}^{\rm U})}$ is the path loss with $d_{k,n}^{\rm U}$ being the distance from RIS$_k$ to UE$_n$, \textcolor{black}{${\mu^{\rm U}_{k,n}}={2\pi d^{\rm U}_{k,n}}/{\lambda}$ is the common phase shift introduced by the LoS distance $d^{\rm U}_{k,n}$}; $\kappa^{\rm{U}}$ represents the Rician factor; $\eta _{k,n,l}^{\rm{U}}\sim {\mathcal{CN}}(0,1)$, $\theta _{k,n,l}^{{\rm{U,D}}}$, and $\phi _{k,n,l}^{{\rm{U,D}}}$ are the complex path gain, vertical and horizontal AoD of path $l$, respectively. Note that for the LoS link indexed by $0$, the range of $\theta _{k,n,0}^{{\rm{U,D}}}$ and $\phi _{k,n,0}^{{\rm{U,D}}}$ can be determined by RIS$_k$ and the coverage area where UE$_n$ is located. Specifically, when UE$_n$ belongs to group $c$, we have $\theta _{k,n,0}^{{\rm{U}},{\rm{D}}} \in [ {\theta _{k,c}^{{\rm{cntr}}} - \delta _{k,c}^{{\rm{cntr}}},\theta _{k,c}^{{\rm{cntr}}} + \delta _{k,c}^{{\rm{cntr}}}} ]$ and $\phi _{k,n,0}^{{\rm{U}},{\rm{D}}} \in [ {\phi _{k,c}^{{\rm{cntr}}} - \delta _{k,c}^{{\rm{cntr}}},\phi _{k,c}^{{\rm{cntr}}} + \delta _{k,c}^{{\rm{cntr}}}} ]$, where ${\delta _{k,c}^{\rm cntr}}$ is the angle spread, ${\theta _{k,c}^{\rm cntr}}$ and ${\phi _{k,c}^{\rm cntr}}$ as the vertical and horizontal AoD from RIS$_k$ to the center of intended coverage of the group $c$, respectively. 



\subsection{Sum Spectral Efficiency}
To support multi-UE communication with reduced overhead for feedback, we adopt the two-stage precoding inspired by the JSDM. The total number of data streams $S=\sum\nolimits_{c = 1}^{C}S_{c}\le \min(N,M^{\rm B})$, where $S_c$ denotes the number of data streams transmitted to UEs in group $c$. The transmit signal ${\bf{x}} = {\bf{FPd}} = \sum\nolimits_{c = 1}^C {{{\bf{F}}_c}{{\bf{P}}_c}{{\bf{d}}_c}}$, where  ${\bf{F}} = [ {{{\bf{F}}_1},{{\bf{F}}_2}, \dots ,{{\bf{F}}_C}} ] \in {{\mathbb C}^{{M^{\rm{B}}} \times S}}$ with ${\bf F}_c\in {\mathbb{C}}^{M^{\rm B}\times S_c}$ is the S-CSI-determined PBF matrix, ${\bf{P}} = {\rm{diag}}\left( {{{\bf{P}}_1},{{\bf{P}}_2}, \dots ,{{\bf{P}}_C}} \right) \in {{\mathbb C}^{S \times S}}$ is the precoding matrix designed with I-CSI, and ${\bf{d}} = {[ {{\bf{d}}_1^T,{\bf{d}}_2^T, \dots ,{\bf{d}}_C^T} ]^T} \in {{\mathbb{C}}^S}$ is the transmit data stream. The PBF matrix ${\bf F}$ transforms the high-dimensional channel ${\bf H}^H \in {\mathbb{C}}^{N \times M^{\rm B}}$ into a reduced-dimension effective channel ${\bf H}^H{\bf F} \in {\mathbb{C}}^{N \times S}$, and ${\bf P}_c\in {\mathbb{C}}^{{S_c}\times {S_c}}$ in ${\bf P}$ is utilized to address the inter-user interference in group $c$. The transmit power constraint is given by ${\mathbb E}\{\|{\bf x}\|^2\}\le P_{\rm max}$.

\textcolor{black}{When $S$ out of $N$ UEs are served, the received signal of UE$_s$ is given by
	\begin{equation}
		{y_s} = {\bf{h}}_s^H{\bf{FPd}} + {z_s}.
	\end{equation}
For ${\bf d}\sim {\mathcal{CN}}(0,{\bf I}_S)$, the downlink sum SE is given by
\begin{equation}
	R = \sum\limits_{s = 1}^S {{\log }_2}{\left( {1 + \frac{{{{\left| {{{\left[ {{{\bf{h}}_s^H}{\bf{FP}}} \right]}_s}} \right|}^2}}}{{\sum\limits_{m = 1,m \ne n}^S {{{\left| {{{\left[ {{{\bf{h}}_s^H}{\bf{FP}}} \right]}_m}} \right|}^2}}  + \sigma _s^2}}} \right)},
\end{equation}
where $[{{{\bf{h}}_s^H}{\bf{FP}}}]_s$ denotes the $s$-th element of ${{{\bf{h}}_s^H}{\bf{FP}}}$.}
The sum SE scales linearly with the number of data streams $S$. The maximum number of data streams that a wireless system can transmit is determined by the rank of the wireless channel. In systems without RISs, the uncontrollable channel necessitates passive adjustments to the number of data streams in response to channel variations. Applying conventional JSDM \cite{JSDM} in this scenario, $S$ is equal to the sum of the principal eigenvalues in the channel covariance matrix for each group. However, the prerequisite is that the eigenmatrices of the channel covariance matrix for each group should be orthogonal to each other. Due to the uncertainties in the propagation environment and UE distributions, satisfying this orthogonality necessitates large-scale antennas at the BS. \textcolor{black}{In certain specific scenarios where UEs are distributed in a line relative to the BS, JSDM fails because the BS cannot differentiate between UE groups or individual UEs based on physical angle information.} 

In wireless systems assisted by RISs, although the direct BS--UE channel remains uncontrollable, the cascaded BS-RISs--UE channel can be modified by manipulating the propagation of electromagnetic waves with RISs. Specifically, when the direct BS--UE channel is severely damaged, RISs achieve the maximum degree of freedom for the end-to-end BS-UE channel, enabling arbitrary channel rank customization \cite{CC-mmwave} and adaptive transmissions \cite{CC-JSAC}. \textcolor{black}{With the assistance of RISs, JSDM can be implemented in a novel manner, even addressing scenarios where traditional JSDM fails. For instance, when all UEs are distributed along the same angular direction from the BS, RISs enable effective separation. In this context, we proceed to investigate, under the worst-case scenario where the direct BS--UE channel is obstructed, how to utilize RISs to implement low-complexity JSDM and enhance the sum SE of UEs in the coverage hole.}

\section{JSDM with RISs-Customized Channel}\label{sec:3}
\textcolor{black}{In this section, we present a multi-RIS-assisted JSDM scheme with a customized channel, which includes channel rank enhancement, inter-group interference suppression, and intra-group interference suppression. Considering that the spatial multiplexing gain, which is essential for supporting multi-UE transmission in JSDM, is directly related to the channel rank between the BS and UEs, we first optimize the positions of RISs to maximize the channel rank in Sec. \ref{sec:3.1}. Based on this deployment, in Sec. \ref{sec:3.2}, the effective channel is block-diagonalized by minimizing inter-group interference that arises from the BS--RISs and RISs--UEs with the PBF design and RIS reflection optimization, respectively. Differing from \cite{JSDM-1RIS} and \cite{JSDM-ICC} that design PBF for the BD through complex matrix decomposition, we construct the PBF matrix with DFT vectors to completely remove the interference of LoS links in the BS--RISs channel, significantly reducing the implementation in terms of both computational complexity and hardware cost. The interference from RISs-UEs channel is reduced by the RIS reflection design in three regimes. After that, in Sec. \ref{sec:3.3}, intra-group interference is dispelled with zero-forcing (ZF) precoding at the BS.}
\subsection{Larger Channel Rank with Position Placement}\label{sec:3.1}
To demonstrate the full potential of RISs in altering the channel rank, we investigate the placement of RISs under the worst condition where the direct BS--UE channel is weak enough to be omitted\footnote{For instance, the direct link is obstructed by a building with at least three concrete walls, each contributing a penetration loss of more than 25dB \cite{Pene-loss}.  \textcolor{black}{As shown in Fig. \ref{Fig.V-system}, six RISs are introduced to restore communication services in a system operating at $6.5$ GHz, where the BS, RIS$_1$, RIS$_2$, and UE$_1$ are located in $[0,0,30]$, $[57,20,37]$, $[79,9,39]$, and $[62,1,0.5]$, respectively. In this case, the large-scale attenuation of the BS--UE$_1$, BS--RIS$_1$--UE$_1$, BS--RIS$_2$--UE$_1$, BS--RIS$_1$--RIS$_2$--UE$_1$, and BS--RIS$_2$--RIS$_1$--UE$_1$ channels are about $-117.7$ dB, $-82.7$ dB, $-84.1$ dB, $-121.1$ dB, and $-120.2$ dB respectively. Compared to the single-cascaded BS--RIS$_k$--UE channel, the direct BS--UE channel and the multi-cascaded BS--RIS$_k$--RIS$_m$--UE channel can be disregarded.}}. Denoting ${{\bf{U}}_k} = [ {{{\bf{u}}_{k,1}},{{\bf{u}}_{k,2}}, \dots ,{{\bf{u}}_{k,N}}} ]\textcolor{black}{\in {\mathbb{C}}^{M^{\rm R}\times N}}$ as RIS$_k$--UEs channel, the rank of the end-to-end channel can be expressed as
\begin{equation}\label{rank-H}
	{\rm{rank}}\left( {\bf{H}} \right) \approx {\rm{rank}}\left( {\sum\limits_{k = 1}^K {{{\bf{B}}_k} {{{\bm{\Gamma}}_k}}{{\bf{U}}_k}} } \right).
\end{equation}
Considering that the BS and RISs are typically deployed on high buildings, the BS--RISs channel exhibits a large Rician factor and is dominated by the LoS channel\footnote{Although the LoS component does not exactly represent the BS--RIS$_k$ channel, it captures the primary characteristics of the channel. In Section \ref{sec:5}, we will demonstrate that NLoS paths with a smaller Rician factor have limited impacts on the performance of the proposed scheme.}, represented as ${{\bf{B}}_k} \approx \beta _{k,0}^{\rm{B}}{{\bf{b}}_{k,0}}{\bf{r}}_{k,0}^H$. In this scenario, \eqref{rank-H} can be upper-bounded as
	\begin{align}
		{\rm{rank}}\left( {\bf{H}} \right) &\approx {\rm{rank}}\left( {\sum\limits_{k = 1}^K {\beta _{k,0}^{\rm{B}}{{\bf{b}}_{k,0}}{\bf{r}}_{k,0}^H {{{\bm{\Gamma}}_k}}{{\bf{U}}_k}} } \right) \notag \\
		& \le \sum\limits_{k = 1}^K {{\rm{rank}}\left( {\beta _{k,0}^{\rm{B}}{{\bf{b}}_{k,0}}{\bf{r}}_{k,0}^H {{{\bm{\Gamma}}_k}}{{\bf{U}}_k}} \right)} ,\label{rank-H-up}
	\end{align}
where the inequality arises because ${\rm rank}(\sum\nolimits_{k = 1}^{K}{\bf A}_k)\le \sum\nolimits_{k= 1}^{K}{\rm rank}({\bf A}_k)$. When the cascaded channels provided by RIS$_k$ and RIS$_m$ ($m \ne k$) satisfy
\begin{equation}\label{rank-H-or}
	\begin{aligned}
		{{{\left( {\beta _{k,0}^{\rm{B}}{{\bf{b}}_{k,0}}{\bf{r}}_{k,0}^H{{{\bm{\Gamma}}_k}} {{\bf{U}}_k}} \right)}}^H}{\left( {\beta _{m,0}^{\rm{B}}{{\bf{b}}_{m,0}}{\bf{r}}_{m,0}^H{{{{\bm{\Gamma}}_m}} }{{\bf{U}}_m}} \right)}= {\bf{0}},
	\end{aligned}
\end{equation}
the upper bound in \eqref{rank-H-up} can be achieved. One implementation of \eqref{rank-H-or} is to install RISs in DFT directions of the BS, which ensures ${\bf{b}}_{k,0}^H{{\bf{b}}_{m,0}} = 0$ for any  $m\ne k$ (the proof is provided in Appendix \ref{App:A0}). \textcolor{black}{This implementation requires perfect alignment between RISs and DFT directions of the BS to achieve strict orthogonality among LoS links. The deviation in the RIS position disrupts this orthogonality and will be evaluated in Sec. \ref{sec:5.D}.} Given that $1 \le {\rm{rank}}( {\beta _{k,0}^{\rm{B}}{{\bf{b}}_{k,0}}{\bf{r}}_{k,0}^H {{{\bm{\Gamma}}_k}} ){{\bf{U}}_k}}  \le {\rm{rank}}( {{{\bf{b}}_{k,0}}} ) = 1$, the maximum rank can be approximated as ${\rm{rank}}( {\bf{H}} ) \approx K$. \textcolor{black}{Although placing RISs at the DFT directions of the BS limits the flexibility of RIS deployment and prevents achieving the optimal overall system link quality, such as ensuring maximal signal-to-noise ratio (SNR), this approach provides the maximum spatial multiplexing gain in the considered scenario. This gain is critical for JSDM schemes that prioritize multi-UE transmission. In general multi-RIS-assisted systems, a key challenge for future research will be to explore how to balance the tradeoff between SNR gain and spatial multiplexing gain through optimal RIS deployment.} 

\subsection{Block-Diagonalized Effective Channel with \textcolor{black}{Pre-Beamforming and Phase Shift Design}}\label{sec:3.2}
\textcolor{black}{The optimized position of RISs provides a channel with preferred spatial multiplexing gain for JSDM. Block-diagonalizing the effective channel is a specific manifestation of how JSDM utilizes spatial multiplexing to reduce inter-group interference.} With the PBF matrix ${\bf F}$ determined by S-CSI, the effective channel is defined as
{ \begin{equation}\label{eq-H-eff-0}
		{{\bf{H}}^H}{\bf{F}} = \left[ {\begin{array}{*{20}{c}}
				{{\bf{H}}_1^H{{\bf{F}}_1}}&{{\bf{H}}_1^H{{\bf{F}}_2}}& \cdots &{{\bf{H}}_1^H{{\bf{F}}_C}}\\
				{{\bf{H}}_2^H{{\bf{F}}_1}}&{{\bf{H}}_2^H{{\bf{F}}_2}}& \cdots &{{\bf{H}}_2^H{{\bf{F}}_C}}\\
				\vdots & \vdots & \ddots & \vdots \\
				{{\bf{H}}_C^H{{\bf{F}}_1}}&{{\bf{H}}_C^H{{\bf{F}}_2}}& \cdots &{{\bf{H}}_C^H{{\bf{F}}_C}}
		\end{array}} \right].
\end{equation}}
In conventional JSDM \cite{JSDM}, ${{\bf{H}}^H}{\bf{F}}$ is block-diagonalized using ${\bf F}$ derived from the channel covariances of UEs in each group. \textcolor{black}{The design of ${\bf F}$, ensuring ${\bf{H}}_{e}^H{{\bf{F}}_c}\approx {\bf 0}$ for all $e\ne c$, involves a complex process including: 1) calculating eigenmatrices of channel covariances; 2) optimizing the number of dominant eigenmodes, the dimensions of ${\bf F}_c$, and the number of data streams in group $c$; and 3) performing two singular value decomposition (SVD) calculations. The challenges in this BD arise from the dynamic nature of the uncontrollable wireless channel. With the introduction of RISs in the wireless channel, we develop a low-complexity approach that designs the PBF and RIS reflection utilizing the angular information of propagation paths.}

Since $K$ orthogonal BS--RIS subchannels are ensured by the placement of $K$ RISs, the data stream number $S$ is set to $K$. Specifically, in group $c$, we have $S_c=K_c$. \textcolor{black}{Note that the subsequent analysis in this section proceeds under the premise of predetermined UE--RIS grouping and association. In Sec. \ref{sec:4}, we will systematically develop optimized grouping and association strategies guided by the principle of channel correlation minimization.} Substituting ${\bf{F}} = [ {{{\bf{f}}_1},{{\bf{f}}_2}, \dots ,{{\bf{f}}_K}} ] $ into the effective channel, we have
		\begin{equation}
			{{\bf{H}}^H}{\bf{F}} ={ \left[ {\sum\limits_{k = 1}^K {{\bf{U}}_k^H{{\bm{\Gamma }}_k^H} {\bf{B}}_k^H} {{\bf{f}}_1}, \dots, \sum\limits_{k = 1}^K {{\bf{U}}_k^H {{\bm{\Gamma }}_k^H} {\bf{B}}_k^H} {{\bf{f}}_K}} \right]}.
            \label{eq-H-eff-1}
		\end{equation}
Considering that the dominant LoS propagation paths of the BS--RISs channel align with the DFT direction of the BS antenna array, we design PBF vector ${{\bf{f}}_i} = {{\bf{b}}_{i,0}}$ to steer transmit energy towards the intended RIS and minimize interference to other RISs. This PBF design, achievable in the analog antenna architecture, significantly reduces the implementation complexity and cost,  as it does not require high-dimensional matrix decomposition and a fully-digital antenna architecture like conventional JSDM schemes. Given that ${{\bf{B}}_k} \approx \beta _{k,0}^{\rm{B}}{{\bf{b}}_{k,0}}{\bf{r}}_{k,0}^H$ and ${\bf{b}}_{k,0}^H{{\bf{b}}_{k',0}}=0$ ($k\ne k'$), the effective channel can be approximated as
\begin{equation}\label{eq-H-eff-2}
			{{\bf{H}}^H}{\bf{F}} \approx {\left[ {\beta _{1,0}^{\rm{B}}{\bf{U}}_1^H{\bm{\Gamma}}_1^H{{\bf{r}}_{1,0}}, \dots ,\beta _{K,0}^{\rm{B}}{\bf{U}}_K^H{\bm{\Gamma}}_K^H{{\bf{r}}_{K,0}}} \right]}.
\end{equation}
By expanding ${\bf U}_k$ and utilizing ${\bf u}^H_{k,n}{\bf \Gamma}^H_{k}{\bf r}_{k,0} =({\bf u}^H_{k,n}\odot {\bf r}^T_{k,0}){\bm \gamma}^*_k $ for \eqref{eq-H-eff-2}, we can express the block matrix ${\bf{H}}_{e}^H{{\bf{F}}_c} \in {{\mathbb{C}}^{{N_{e}} \times {K_c}}}$ as shown in \eqref{eq-H-eff-3}, where subscripts $k_j\in{\mathcal K}_c$ for $j=1,\dots,K_c$ and $n_j\in {\mathcal N}_e$ for $j=1,\dots,N_{e}$. \textcolor{black}{The channel approximation in \eqref{eq-H-eff-2} indicates that, on the basis of the optimized RIS placement derived in Sec. \ref{sec:3.1}, using DFT codewords to construct the PBF effectively reduces the interference between the BS-RIS subchannels.  Subsequently, we will discuss how to combine this PBF design with the optimization of phase shifts of the RIS to reduce the inter-group interference between RISs-UEs subchannels, thereby block-diagonalizing the effective channel.}

\begin{figure*}
	{ \begin{equation}\label{eq-H-eff-3}
		{\bf{H}}_{e}^H{{\bf{F}}_c} = \left[ {\left[ {\begin{array}{*{20}{c}}
					{\beta _{k_{1},0}^{\rm{B}}\left( {{\bf{u}}_{k_{1},n_{1}}^H \odot {\bf{r}}_{k_{1},0}^T} \right)}\\
					\vdots \\
					{\beta _{k_{1},0}^{\rm{B}}\left( {{\bf{u}}_{k_{1},n_{N_e}}^H \odot {\bf{r}}_{k_{1},0}^T} \right)}
			\end{array}} \right]{\bm{\gamma }}_{k_{1}}^*, \dots ,\left[ {\begin{array}{*{20}{c}}
					{\beta _{k_{K_c},0}^{\rm{B}}\left( {{\bf{u}}_{k_{K_c},n_{1}}^H \odot {\bf{r}}_{k_{K_c},0}^T} \right)}\\
					\vdots \\
					{\beta _{k_{K_c},0}^{\rm{B}}\left( {{\bf{u}}_{k_{K_c},n_{N_e}}^H \odot {\bf{r}}_{k_{K_c},0}^T} \right)}
			\end{array}} \right]{\bm{\gamma }}_{k_{K_c}}^*} \right]
	\end{equation}}
\hrulefill
\end{figure*}

\textcolor{black}{Considering $e = c$ for \eqref{eq-H-eff-3},} we design reflection vectors of RISs in ${\mathcal K}_c$ to concentrate transmit energy on UEs in group $c$. Specifically, the reflection vector ${{\bm{\gamma }}_k}$ ($k\in{\mathcal K}_c$)  is designed\footnote{\textcolor{black}{The reflection vectors of RISs are designed based on the direction of channel paths. The coherence time of this statistical CSI can be an order of magnitude longer than the channel coherence time, even in mobility scenarios \cite{coherence}. Given the reciprocity of statistical CSI in FDD systems \cite{Reci}, the BS can readily obtain these path directions for configuring RISs without feedback from UEs, using a reliable fixed control channel. For analytical tractability, we assume that the reflection vector of RISs is continuous and perfectly configured. In Sec. \ref{sec:5.D}, we will evaluate the impact of practical impairments, such as phase noise, deployment offsets, and discrete phases of RIS elements, on the performance of the proposed JSDM.}} as
\begin{equation}\label{eq-Gamma}
	{{\bm{\gamma }}_k} = {M^{\rm R}}\left( {{{\bf{a}}}\left( {{\Delta ^{\rm{v}}_{k}},{\Delta ^{\rm{h}}_{k}}} \right) \odot {{\bf{r}}^{*}_{k,0}}} \right)^{*},
\end{equation}
where ${\bf a}(\cdot,\cdot)$ and ${\bf r}_{k,0}$ are normalized ARVs, $M^{\rm R}$ ensures unit amplitude for elements in ${\bm \gamma}_k$; \textcolor{black}{${\Delta ^{\rm{v}}_{k}} = \theta _{k,c}^{\rm cntr} + {\delta^{\rm bias} _{{\rm{v}},k}}$ and ${\Delta ^{\rm{h}}_{k}} = \phi _{k,c}^{\rm cntr} + {\delta^{\rm bias} _{{\rm{h}},k}}$ are the vertical and horizontal reflection directions, respectively; ${\delta^{\rm bias} _{{\rm{v}},k}}$ and ${\delta^{\rm bias} _{{\rm{h}},k}}$ are beam biases limited in the range $ [ { - \delta _{k,c}^{\rm cntr},\delta _{k,c}^{\rm cntr}} ]$. The design in \eqref{eq-Gamma} redirects the signal from the LoS link of the BS--RIS$_k$ channel around the center of coverage represented by $(\theta _{k,c}^{\rm cntr},\phi _{k,c}^{\rm cntr})$ with beam biases ${\delta^{\rm bias} _{{\rm{v}},k}}$ and ${\delta^{\rm bias} _{{\rm{h}},k}}$. The following analysis will design these beams biases to enhance the BD of the effective channel.}

Customizing exact orthogonal group channels such that ${\bf{H}}_{e}^H{{\bf{F}}_c} = {\bf{0}}$ for $e \ne c$ is challenging using only directional parameters. The approximated BD can be approached when the ratio between the interference from UEs in other groups and the received signal of the intended UE approaches zero. For the RIS$_k$ assigned to UE$_n$ within group $c$, the condition can be formulated as follows:
\begin{equation}\label{eq-ratio-1}
	\frac{{\left| {{{[{\bf{H}}_e^H{{\bf{F}}_c}]}_{n',k}}} \right|}}{{\left| {{{[{\bf{H}}_c^H{{\bf{F}}_c}]}_{n,k}}} \right|}}=\frac{\left|{{{ { \left( {{\bf{u}}_{k,n'}^H \odot {\bf{r}}_{k,0}^T} \right){\bm{\gamma }}_{k}^*} }}}\right|}{\left|{{{ {\left( {{\bf{u}}_{k,n}^H \odot {\bf{r}}_{k,0}^T} \right){\bm{\gamma }}_{k}^*} }}}\right|} \to 0,
\end{equation}
for $\forall n' \in {\mathcal N}\backslash {\mathcal N}_c$ and $n \in {{\mathcal N}_c}$, where $[{\bf A}]_{n,k}$ denotes the element in the $n$-th row and $k$-the column of ${\bf A}$.

Considering that ${\bf H}^H{\bf F}$ is reshaped with the S-CSI, we provide its statistical property in the following proposition:
\begin{proposition}\label{Pro-1}
    With the PBF aligned with the dominant LoS paths of the BS--RISs channel, and the RIS reflection vector designed in accordance with \eqref{eq-Gamma}, the mean of the element in the $n$-th row and $k$-th column of ${\bf H}^H{\bf F}$ is given by
	\begin{equation}\label{eq-Pro-1}
		\begin{aligned}
			&{\mathbb E}{\left\{ \left[{\bf H}^H{\bf F}\right]_{n,k} \right\}}={\mathbb E}{\left\{ {\beta _{k,0}^{\rm{B}}\left( {{\bf{u}}_{k,n}^H \odot {\bf{r}}_{k,0}^T} \right){\bm{\gamma }}_k^*} \right\}} \\
			&=\beta _{k,0}^{\rm{B}}\beta _{k,n,0}^{\rm{U}}	{\mathcal D}\left(  {\Theta _{k,n}},{\Phi _{k,n}} \right) {e^{j\left( {{\Theta _{k,n}}\left( {{M^{{\rm{R,v}}}} - 1} \right) + {\Phi _{k,n}}\left( {{M^{{\rm{R,h}}}} - 1} \right)} \right)}},
		\end{aligned}
	\end{equation}
where ${\mathcal D}(  {\Theta _{k,n}},{\Phi _{k,n}} )$ is given by
\begin{equation}
	{\mathcal D}\left(  {\Theta _{k,n}},{\Phi _{k,n}} \right) = \frac{{\sin \left( {{M^{{\rm{R,v}}}}{\Theta _{k,n}}} \right)\sin \left( {{M^{{\rm{R,h}}}}{\Phi _{k,n}}} \right)}}{{\sin \left( {{\Theta _{k,n}}} \right)\sin \left( {{\Phi _{k,n}}} \right)}},
\end{equation}
with ${\Theta _{k,n}} = \frac{\pi }{2}\cos ( {{\Delta ^{{\rm{v}}}_k}} ) - \frac{\pi }{2}\cos ( {\theta _{k,n,0}^{{\rm{U,D}}}} )$, and ${\Phi _{k,n}} = \frac{\pi }{2}\sin ( {{\Delta ^{{\rm{v}}}_k}} )\sin ( {{\Delta ^{{\rm{h}}}_k}} ) - \frac{\pi }{2}\sin ( {\theta _{k,n,0}^{{\rm{U,D}}}} )\sin ( {\phi _{k,n,0}^{{\rm{U,D}}}} )$.
\end{proposition}
\begin{IEEEproof}
	Please refer to Appendix \ref{App:A}.
\end{IEEEproof}

Proposition \ref{Pro-1} provides a foundational understanding that ${\mathbb E}\{ [{\bf H}^H{\bf F}]_{n,k} \}$ is exclusively determined by the parameters of the LoS path in the BS--RIS$_k$ and RIS$_k$--UE$_n$ channels. When the Rician factors are sufficiently large, it can be inferred that $ [{\bf H}^H{\bf F}]_{n,k}  \to {\mathbb E}\{ [{\bf H}^H{\bf F}]_{n,k} \} $ holds because $ {\bf H}^H{\bf F}$ is dominated by LoS components, a phenomenon also known as channel hardening \cite{hardening}. \textcolor{black}{By leveraging Proposition \ref{Pro-1}, the focus of the approximated BD shifts from \eqref{eq-ratio-1} to its statistical version:
\begin{equation}\label{eq-abs-ratio-1}
	 {\rm ISR}_{k,n,n'} \buildrel \Delta \over =\frac{{\left|{\mathbb E} {\left\{ {\left( {{\bf{u}}_{k,n'}^H \odot {\bf{r}}_{k,0}^T} \right){\bm{\gamma}}_{k}^*} \right\}} \right|}}{{\left|{\mathbb E} {\left\{ {\left( {{\bf{u}}_{k,n}^H \odot {\bf{r}}_{k,0}^T} \right){\bm{\gamma}}_{k}^*} \right\}} \right|}} \to 0,
\end{equation}
where ${\rm ISR}_{k,n,n'}$ can be regarded as the inter-group interference-to-signal ratio between UE$_{n'}$ ($n'\notin {\mathcal N}_c$) and UE$_n$ ($n\in {\mathcal N}_c$) assisted by RIS$_k$.} This ratio is closely related to the array size of the RIS and the reflected angle biases ${\delta^{\rm bias} _{{\rm{v}},k}}$ and ${\delta^{\rm bias} _{{\rm{h}},k}}$. Therefore, subsequent analysis will separately examine the design of these angle biases for the approximated BD in three cases.

\subsubsection{Asymptotic Orthogonal Regime with Large-Size RIS}\label{sec:3.2.1}
In this scenario, we assume that RISs are sufficiently large to ensure asymptotic orthogonality for ARVs. Following the analysis process detailed in Appendix \ref{App:A},  the numerator in \eqref{eq-abs-ratio-1}  can be calculated as
\begin{equation}\label{eq-3.2.1-inter}
	\begin{aligned}
		\textcolor{black}{{{\left|{\mathbb E} {\left\{ {\left( {{\bf{u}}_{k,n'}^H \odot {\bf{r}}_{k,0}^T} \right){\bm{\gamma}}_{k}^*} \right\}} \right|}} = \left| \beta _{k,n',0}^{\rm{U}}\right|{\mathcal D}\left(  {{\Theta _{k,n'}}},{{\Phi _{k,n'}}} \right)},
	\end{aligned}
\end{equation}
where ${\Theta _{k,n'}} = \frac{\pi }{2}\cos ({\Delta ^{\rm{v}}_{k}}) - \frac{\pi }{2}\cos (\theta _{k,n',0}^{{\rm{U}},{\rm{D}}})$ and ${\Phi _{k,n'}} = \frac{\pi }{2}\sin ({\Delta ^{\rm{v}}_{k}})\sin ({\Delta ^{\rm{h}}_{k}}) - \frac{\pi }{2}\sin (\theta _{k,n',0}^{{\rm{U}},{\rm{D}}})\sin (\phi _{k,n',0}^{{\rm{U}},{\rm{D}}})$. \textcolor{black}{According to the definition below \eqref{eq-Gamma}, for the RIS$_k$ assigned to UE$_n$ within group $c$, $ {\Delta ^{\rm{v}}_{{k}}}\in [ {\theta _{k,c}^{\rm cntr} - \delta _{k,c}^{\rm cntr},\theta _{k,c}^{\rm cntr}+\delta _{k,c}^{\rm cntr}} ]$ and ${\Delta ^{\rm{h}}_{{k}}}\in  [ {\phi _{k,c}^{\rm cntr}  - \delta _{k,c}^{\rm cntr},\phi _{k,c}^{\rm cntr} +\delta _{k,c}^{\rm cntr}} ]$.  Given that UE$_{n'}$ is located outside group $c$, achieving ${\Delta ^{\rm{v}}_{k}}=\theta _{k,n',0}^{{\rm{U}},{\rm{D}}} $ and ${\Delta ^{\rm{h}}_{k}}=\phi _{k,n',0}^{{\rm{U}},{\rm{D}}} $ is impossible. Thus,  ${\Theta _{k,n'}} $ and ${\Phi _{k,n'}} $ remain non-zero constants.} With a sufficiently large RIS, defined by large values of  $M^{\rm R,v}$ and $M^{\rm R,h}$, we have
\begin{equation}
{\mathcal D}\left(  {{\Theta _{k,n'}}},{{\Phi _{k,n'}}} \right)\to 0,
\end{equation}
\textcolor{black}{which satisfies \eqref{eq-ratio-1} for BD in a statistical way.} Further adjustments such as setting $\delta _{{\rm{v}},{k}}^{{\rm{bias}}} =\theta _{k,n,0}^{{\rm{U}},{\rm{D}}} -\theta _{{k},c}^{\rm cntr}$ and $\delta _{{\rm{h}},{k}}^{{\rm{bias}}} =\phi _{k,n,0}^{{\rm{U}},{\rm{D}}} -\phi _{{k},c}^{\rm cntr}$, yield ${\Theta _{k,n}} = 0$ and ${\Phi _{k,n}} = 0$, thus maximizing the denominator in \eqref{eq-abs-ratio-1} to \textcolor{black}{$M^{\rm R}|\beta _{k,n,0}^{\rm{U}}|$}.

In scenarios where asymptotic orthogonality is not feasible, optimizing the biases $({\delta^{\rm bias} _{{\rm{v}},k}},{\delta^{\rm bias} _{{\rm{h}},k}})$ in interference-limited and noise-limited regimes separately allow for the customization of an approximate block-diagonal channel.

\subsubsection{ Interference-limited Regime with Common-Size RIS}\label{sec:3.2.2}
\textcolor{black}{In the interference-limited regime, where noise levels are sufficiently low to be disregarded, $ {{{[{\bf{H}}_e^H{{\bf{F}}_c}]}_{n',k}}} $, which can be considered as interference from out-of-group UEs, becomes the critical factor affecting SE of UEs in group $c$. Therefore, we focus on optimizing the reflection beam biases to minimize the maximal ${\rm ISR}_{k,n,n'}$ given in \eqref{eq-abs-ratio-1}. Following the analysis process detailed in Appendix \ref{App:A} to unfolding ${\rm ISR}_{k,n,n'}$, this optimization problem is formulated as follows:}
\begin{equation}\label{eq-opti-IL}
	\begin{aligned}
	\mathop {\min }\limits_{\delta _{{\rm{v}},k}^{{\rm{bias}}},\delta _{{\rm{h}},k}^{{\rm{bias}}}}
    &\mathop {\max }\limits_{n' \in {\mathcal N}\backslash {{\mathcal {N}}}_c,n\in  {{\mathcal {N}}}_c} \textcolor{black}{\left|\frac{{\beta _{k,n',0}^{\rm{U}}}}{{\beta _{k,n,0}^{\rm{U}}}}\right|}\frac{{\mathcal D}\left(  {{\Theta _{k,n'}}},{{\Phi _{k,n'}}} \right)}{{\mathcal D}\left(  {{\Theta _{k,n}}},{{\Phi _{k,n}}} \right)}\\
		{\rm{s.t.}} & \qquad \delta _{{\rm{v}},k}^{{\rm{bias}}},\delta _{{\rm{h}},k}^{{\rm{bias}}} \in \left[ { - \delta _{k,c}^{\rm cntr},\delta _{k,c}^{\rm cntr}} \right].
	\end{aligned}
\end{equation}
Given the constrained optimization region, \eqref{eq-opti-IL} is tackled through an exhaustive search with steps of $\delta _{{\rm{v}},k}^{{\rm{bias}}}$ and $\delta _{{\rm{h}},k}^{{\rm{bias}}}$ set as ${2\pi}/{M^{\rm R,v}}$ and  ${2\pi}/{M^{\rm R,h}}$, respectively.

\subsubsection{Noise-Limited Regime with Common-Size RIS}\label{sec:3.2.3}
While approximated BD is crucial for enhancing the SE by reducing inter-group interference, it is essential to preserve sufficient channel power within the intended group of UEs in a noise-limited regime. Unlike the approach in \eqref{eq-opti-IL}, which minimizes the maximal ${\rm ISR}_{k,n,n'}$ without considering the strength of the received signal, in the noise-limited regime, we aim to \textcolor{black}{minimize the maximal interference (the numerator of ${\rm ISR}_{k,n,n'}$) while ensuring that the received signal strength (the denominator of ${\rm ISR}_{k,n,n'}$) remains satisfactory.} This objective is formulated as
\begin{equation}\label{eq-opti-NL}
	\begin{aligned}
		\mathop {\min }\limits_{\delta _{{\rm{v}},k}^{{\rm{bias}}},\delta _{{\rm{h}},k}^{{\rm{bias}}}} &\mathop {\max }\limits_{n' \in {\mathcal N}\backslash {{\mathcal {N}}}_c,n\in  {{\mathcal {N}}}_c} {{\mathcal D}\left(  {{\Theta _{k,n'}}},{{\Phi _{k,n'}}} \right)}\\
		{\rm{s.t.}} \qquad & {{\mathcal D}\left(  {{\Theta _{k,n}}},{{\Phi _{k,n}}} \right)} \ge \tau {M^{\rm{R}}},\\
		&  \delta _{{\rm{v}},k}^{{\rm{bias}}},\delta _{{\rm{h}},k}^{{\rm{bias}}} \in \left[ { - \delta _{k,c}^{\rm cntr},\delta _{k,c}^{\rm cntr}} \right].
	\end{aligned}
\end{equation}
Here, $0\le \tau \le 1$  represents a threshold, typically set to $1/\sqrt{2}$ to ensure that UE$_n$ falls within the half-power beamwidth of the RIS$_{k}$.

\textcolor{black}{Block-diagonalizing ${\bf H}^H{\bf F}$ by designing beam biases ${\delta^{\rm bias} _{{\rm{v}},k}}$ and ${\delta^{\rm bias} _{{\rm{h}},k}}$	in the three cases outlined above represents a tradeoff in controlling the signal-to-interference-plus-noise ratio (SINR). In the first case, where the RIS provides a sufficiently large array gain, the desired signal dominates the SINR, and the beam biases are optimized to maximize the desired signal. In the interference-limited case, solving the optimization problem in \eqref{eq-opti-IL}  improves the signal-to-interference ratio for UEs. Finally, in the case where both interference and noise are significant, we optimize the beam biases of the RISs to minimize interference while maintaining an acceptable SNR.}

\subsection{Intra-Group Interference Elimination with Zero-Forcing Precoding}\label{sec:3.3}

Based on the approximated block-diagonal ${\bf H}^H{\bf F}$, where ${{\bf{H}}^H_c}{{\bf{F}}_{e}}\approx {\bf 0}$ for $c\ne e$, the received signal of UEs in group $c$ can be expressed as
	\begin{equation}
	{{\bf{y}}_c} = {{\bf{H}}}^H_c\sum\limits_{e = 1}^C {{{\bf{F}}_e}{{\bf{P}}_e}{{\bf{d}}_e}} + {{\bf{z}}_c} \approx {{\bf{H}}^H_c}{{\bf{F}}_c}{{\bf{P}}_c}{{\bf{d}}_c} + {{\bf{z}}_c}.
\end{equation}
The ZF precoding is then utilized to eliminate intra-group interference. This precoding method is defined as
\begin{equation}
	{{\bf{P}}_c} = {\varepsilon }{{\bf{H}}_c}{\left( {{{\bf{H}}^H_c}{{\bf{H}}_c}} \right)^{ - 1}},
\end{equation}
where $	\varepsilon^2 = {{{P_{\max }}}}/{{\sum\nolimits_{c = 1}^C {{\rm{tr}}{\{ {{{( {{{\bf{H}}^H_c}{{\bf{H}}_c}} )}^{ - 1}}} \}} } }}$ represents a power normalization factor. \textcolor{black}{Since ${\bf F}$ and ${\bf P}$ can be implemented with analog and digital beamforming in a hybrid architecture, the proposed JSDM scheme eliminates the need for an expensive fully-digital antenna array.} 

\subsection{Complexity Analysis}
\textcolor{black}{
The JSDM scheme consists of two matrix design processes: PBF and post-precoding. Since we adopt the same post-precoding technique as the classical JSDM, namely ZF precoding, the primary difference in complexity arises from the design of the PBF. When applying the classical JSDM method to RIS-assisted systems \cite{JSDM-1RIS}, the PBF design requires one eigenvalue decomposition and two SVDs, resulting in a computational complexity of ${\mathcal O}((M^{\rm B})^3)$. In contrast, we directly design the PBF using ARVs, which are determined by the directional information of the LoS paths in the BS--RIS channels, significantly reducing the computational complexity to ${\mathcal O}(1)$.
Moreover, the use of directional information from LoS paths ensures that the PBF does not require frequent updates when the BS and RISs are fixed. These advantages are enabled by the RIS deployment and reflection design. Specifically, for RIS$_k$, we design the reflection vector using a closed-form expression determined by the S-CSI. The reflection design optimization, including the problems in \eqref{eq-opti-IL} and \eqref{eq-opti-NL}, involves ${M^{\rm{R}}}{( {\delta _{k,c}^{{\rm{cntr}}}/\pi } )^2}$ searches, each with a computational complexity of ${\mathcal O}(1)$.
For comparison, the gradient ascent-based reflection design proposed in \cite{JSDM-1RIS} has a complexity of ${\mathcal O}(C(M^{\rm B}(M^{\rm R})^2)+M^{\rm R}+M^{\rm B})$.}

\section{Channel Correlation Reduction with UE-RIS Grouping and Association}\label{sec:4}
\textcolor{black}{In the previous section, we block-diagonalizing the effective channel under the given UE-RIS grouping and association. However, the BD result of ${\bf H}^H{\bf F}$  also depends on the grouping and association of UEs and RISs. For instance, the grouping of UEs/RIS determines the relative position of rows/columns in the effective channel, which forms the basic structure of ${\bf H}^H{\bf F}$. Because the channel rank is subjected to the number of RISs,  when the number of UEs exceeds that of RISs, not all UEs can be provided with independent data streams. } In this case, the RIS-UE association that allocates RIS resources among UEs is necessary to exclude highly correlated UEs and achieve the desired block diagonal structure of ${\bf H}^H{\bf F}$.

\textcolor{black}{In this section, we commence with an analytical characterization of the effective channel correlation between any two UEs. To form $C$ distinct UE-RIS clustering with minimized inter-group channel correlation, we employ the K-means clustering and a path loss method for the UE grouping and RIS grouping, respectively, enhancing the BD performance.} The UE-RIS association within each group is then developed to further reduce the correlation of elements in ${\bf H}^H{\bf F}$. 

\subsection{\textcolor{black}{Effective Channel Correlation Analysis}}
The effective channel cross-correlation between different UEs is influenced by the deployment and reflection design of RISs, as well as the spatial characteristics of the UEs. Building upon the previous section on RIS deployment and phase shift configuration, we investigate the impact of the spatial distribution parameters of UEs on the cross-correlation of the effective channels.

\textcolor{black}{Given that the design of the PBF matrix ${\bf F}$ is determined by the fixed positions of RISs, the effective channel for UE$_n$ can be represented as ${\bf{h}}_n^H{\bf{F}}$. The effective channel cross-correlation between any UE$_n$ and UE$_m$ is then defined as}
\begin{equation}\label{eq-Qnm}
	{Q_{n,m}} = {\mathbb{E}}\left\{ {{\bf{h}}_m^H{\bf{F}}{{\bf{F}}^H}{{\bf{h}}_n}} \right\}.
\end{equation}

\textcolor{black}{\begin{lemma}\label{lemma-Qnm}
	In cases where LoS paths dominate in the BS--RISs channel, assuming that the RIS$_k$ is assigned to the UE$_k$\footnote{The UE--RIS association will be detailed in the subsequent content.}, that is, ${{\bm{\gamma}}_k} = {M^{\rm R}}( {{\bf{r}}_{k,k,0}^*  \odot  {{\bf{r}}_{k,0}}  } )$, the effective channel cross-correlation in \eqref{eq-Qnm} can be expressed as
	\begin{equation}\label{eq-Qnm_app-1}
		{Q_{n,m}} \approx\sum\limits_{k = 1}^K {{\beta _{k,n,m}}\left( {{\bf{r}}_{k,m,0}^H{{\bf{r}}_{k,k,0}}} \right)\left( {{\bf{r}}_{k,k,0}^H{{\bf{r}}_{k,n,0}}} \right)},
	\end{equation}
where ${\beta _{k,n,m}} = {| {\beta _{k,0}^{\rm{B}}} |^2}\beta _{k,n,0}^{\rm{U}}\beta _{k,m,0}^{\rm{U}}$ incorporates parameters of path losses, and ${{\bf r}_{k,n,0}}$ is the ARV of the LoS path from the RIS$_k$ to the UE$_n$.
\end{lemma}
\begin{IEEEproof}
	Please refer to Appendix \ref{App:B}.
\end{IEEEproof}}
\textcolor{black}{\emph{Lemma} \ref{lemma-Qnm}  indicates that $Q_{n,m}$ is related to the direction and path loss (distance) of each LoS path in the RISs--UE$_n$ and RISs--UE$_m$ channels. This statistical characteristic, determined by the reciprocal S-CSI, provides a fundamental principle for the grouping of UEs and RISs.}

\subsection{Grouping of UEs and RISs}\label{sec-4.2}
To achieve a favorable block-diagonal effective channel, the channel cross-correlation between UE$_n$ and UE$_m$ in different groups should be minimized. In other words, UEs with high correlation should be grouped together. Defining ${{\bf{v}}_n} = {[ {\theta _{1,n,0}^{{\rm{U}},{\rm{D}}},\phi _{1,n,0}^{{\rm{U}},{\rm{D}}}, \dots ,\theta _{K,n,0}^{{\rm{U}},{\rm{D}}} ,\phi _{K,n,0}^{{\rm{U}},{\rm{D}}}} ]^T}$ as the direction vector of LoS paths from all RISs to UE$_n$, \textcolor{black}{according to \emph{Lemma} \ref{lemma-Qnm}}, UEs with high correlation will have a small Euclidean distance for their direction vector. Thus, we employ K-means clustering to divide $\{{\bf v}_n\}_{n=1}^{N}$ into $C$ groups. Specifically, we begin with the random selection of $C$ centers for $\{{\bf v}_n\}_{n=1}^{N}$. Each direction vector is then assigned to the nearest center based on the Euclidean distance, forming $C$ clusters. The centers are recalculated as the mean of all direction vectors in their respective clusters. This process of reassignment and center update continues until the change in centers falls below a threshold. With the result of K-means clustering, sets of UEs, $\{{\mathcal N}_c\}^C_{c=1}$, can be obtained.  In the following two paragraphs, intra- and inter-group channel cross-correlations are analyzed respectively to show the effectiveness of this grouping in channel BD.

When UE$_n$ and UE$_m$ belong to the same group $c$, we rewrite \eqref{eq-Qnm_app-1} as
	\begin{align}
		&{Q_{n,m}} \approx  \sum\limits_{k \in {{\mathcal K}_c}} {{\beta _{k,n,m}}\left( {{\bf{r}}_{k,m,0}^H{{\bf{r}}_{k,k,0}}} \right)\left( {{\bf{r}}_{k,k,0}^H{{\bf{r}}_{k,n,0}}} \right)} \notag  \\
		&+ \sum\limits_{k \notin {{\mathcal K}_c}} {{\beta _{k,n,m}}\left( {{\bf{r}}_{k,m,0}^H{{\bf{r}}_{k,k,0}}} \right)\left( {{\bf{r}}_{k,k,0}^H{{\bf{r}}_{k,n,0}}} \right)} .
        \label{eq-Qnm_app-2}
	\end{align}
\textcolor{black}{For the first term of \eqref{eq-Qnm_app-2} where $k\in{\mathcal K}_c$,} the AoDs of group $c$ satisfy
\begin{equation} \label{eq-intra-AoD}
	\left\{ \begin{aligned}
		&| {\theta _{k,k,0}^{{\rm{U}},{\rm{D}}} - \theta _{k,x,0}^{{\rm{U}},{\rm{D}}}} | \le 2\delta _{k,c}^{{\rm{cntr}}}\\
		&| {\phi _{k,k,0}^{{\rm{U}},{\rm{D}}} - \phi _{k,x,0}^{{\rm{U}},{\rm{D}}}} | \le 2\delta _{k,c}^{{\rm{cntr}}}
	\end{aligned} \right.,
\end{equation}
where $x\in\{n,m\}$ and ${\delta _{k,c}^{\rm cntr}}$ is the angle spread of the coverage $c$. With a small angle spread, we have $ {{\bf{r}}_{k,m,0}^H{{\bf{r}}_{k,k,0}}}\to 1$ and ${{\bf{r}}_{k,k,0}^H{{\bf{r}}_{k,n,0}}}\to 1$ for the first term of \eqref{eq-Qnm_app-2}. For $k\in {\mathcal K}_{e}$ ($e\ne c$), we have
\begin{equation}\label{eq-inter-AoD}
	\left\{ \begin{aligned}
		&| {\theta _{k,k,0}^{{\rm{U}},{\rm{D}}} - \theta _{k,x,0}^{{\rm{U}},{\rm{D}}}} | \ge | {\theta _{k,e}^{{\rm{cntr}}} - \theta _{k,c}^{{\rm{cntr}}}} | - \delta _{k,e}^{{\rm{cntr}}} - \delta _{k,c}^{{\rm{cntr}}}\\
		&| {\phi _{k,k,0}^{{\rm{U}},{\rm{D}}} - \phi _{k,x,0}^{{\rm{U}},{\rm{D}}}} | \ge | {\phi _{k,e}^{{\rm{cntr}}} - \phi _{k,c}^{{\rm{cntr}}}} | - \delta _{k,e}^{{\rm{cntr}}} - \delta _{k,c}^{{\rm{cntr}}}
	\end{aligned} \right.,
\end{equation}
where ${\theta _{k,c}^{\rm cntr}}$ and ${\phi _{k,c}^{\rm cntr}}$ are AoDs from RIS$_k$ to the center of coverage $c$. \textcolor{black}{Considering that groups $c$ and $e$ have distinct coverage centers and angle spreads of these coverage areas are small, ${| {\theta _{k,e}^{\rm cntr} - \theta _{k,c}^{\rm cntr}} |} $ and ${| {\phi _{k,e}^{\rm cntr} - \phi _{k,c}^{\rm cntr}} |}$ dominate the right terms of \eqref{eq-inter-AoD} because they are much larger than $\delta _{k,e}^{\rm cntr} + \delta _{k,c}^{\rm cntr}$. In this case, ${| {\theta _{k,k,0}^{{\rm{U}},{\rm{D}}} - \theta _{k,x,0}^{{\rm{U}},{\rm{D}}}} |}$ and ${| {\phi _{k,k,0}^{{\rm{U}},{\rm{D}}} - \phi _{k,x,0}^{{\rm{U}},{\rm{D}}}} |}$  are sufficiently large to ensure $ {{\bf{r}}_{k,m,0}^H{{\bf{r}}_{k,k,0}}}\to 0$ and ${{\bf{r}}_{k,k,0}^H{{\bf{r}}_{k,n,0}}}\to 0$ for the second term of \eqref{eq-Qnm_app-2}.} In other words, $Q_{n,m}$ is dominated by the first term of \eqref{eq-Qnm_app-2} in this case.

In the other case where UE$_n$ and UE$_m$ belong to different groups $c$ and $e$, respectively, \eqref{eq-Qnm_app-1} can be separated as
	\begin{align}\label{eq-Qnm_app-3}
		&{Q_{n,m}} \approx  \sum\limits_{k \in {{\mathcal K}_c}} {{\beta _{k,n,m}}\left( {{\bf{r}}_{k,m,0}^H{{\bf{r}}_{k,k,0}}} \right)\left( {{\bf{r}}_{k,k,0}^H{{\bf{r}}_{k,n,0}}} \right)} \notag \\
		 + &\sum\limits_{k \in {{\mathcal K}_{e}}} {{\beta _{k,n,m}}\left( {{\bf{r}}_{k,m,0}^H{{\bf{r}}_{k,k,0}}} \right)\left( {{\bf{r}}_{k,k,0}^H{{\bf{r}}_{k,n,0}}} \right)} \notag \\
		+ &\sum\limits_{{k \notin {{\mathcal K}_c}}\cup {\mathcal K}_{e}} {{\beta _{k,n,m}}\left( {{\bf{r}}_{k,m,0}^H{{\bf{r}}_{k,k,0}}} \right)\left( {{\bf{r}}_{k,k,0}^H{{\bf{r}}_{k,n,0}}} \right)} .    
	\end{align}
Considering the differences of intra- and inter-group AoDs in \eqref{eq-Qnm_app-3}, we have $ {{\bf{r}}_{k,m,0}^H{{\bf{r}}_{k,k,0}}}\to 0$ and ${{\bf{r}}_{k,k,0}^H{{\bf{r}}_{k,n,0}}}\to 1$ for the first term, $ {{\bf{r}}_{k,m,0}^H{{\bf{r}}_{k,k,0}}}\to 1$ and ${{\bf{r}}_{k,k,0}^H{{\bf{r}}_{k,n,0}}}\to 0$ for the second term, and $ {{\bf{r}}_{k,m,0}^H{{\bf{r}}_{k,k,0}}}\to 0$ and ${{\bf{r}}_{k,k,0}^H{{\bf{r}}_{k,n,0}}}\to 0$ for the third term. The first two terms of \eqref{eq-Qnm_app-3} are much smaller than the first term of \eqref{eq-Qnm_app-2} because of the tiny inner product. This result exhibits a rudiment for BD.

The above analysis demonstrates how the UE grouping affects $Q_{n,m}$ through the inner product terms. \textcolor{black}{As shown in \eqref{eq-Qnm_app-1}, $\beta_{k,n,m}$ is another factor that can change $Q_{n,m}$ for different UEs. UEs in different groups necessitate small values of  $|\beta_{k,n,m}|$ to diminish $Q_{n,m}$. Hence RISs with larger $|\beta_{k,n,m}|$ are assembled into the same group. According to the definition below \eqref{eq-Qnm_app}, we know that $|{\beta _{k,n,m}}| = {| {\beta _{k,0}^{\rm{B}}} |^2} {\frac{{{M^{\rm R}}{\kappa ^{\rm{U}}}}}{{{\kappa ^{\rm{U}}} + 1}}} \rho _{k,n}^{\rm{U}}\rho _{k,m}^{\rm{U}}$} is proportional to the path losses from RIS$_k$ to UE$_n$ and to UE$_m$. Thus, we proposed a path loss-based RISs grouping scheme in Algorithm \ref{alg:RISs-grouping}. The algorithm involves initially calculating the average path loss $\bar \rho _{k,c}^{{\rm{group}}}$ from RIS$_k$ to UEs in group $c\in {\mathcal C}$, where ${\mathcal C}$ is the set of groups. Then, in descending order of $\{\bar \rho _{k,c}^{{\rm{group}}},k\in{\mathcal K},c\in{\mathcal C}\}$,  RISs are allocated to each group in sequence.  Note that the maximum number of RISs for group $c$ is $|{\mathcal N}_c|$.

\begin{algorithm}[htb]
	\caption{Path Loss-based RISs Grouping}

	\hspace*{0.02in} {\bf Input:} Sets of RISs and UEs: ${\mathcal K}$ and ${\mathcal N}=\{{\mathcal N}_c| c=1,\dots,C\}$; Path loss in the RISs--UEs channel: $\{\rho_{k,n}^{\rm U}| k\in {\mathcal K},n\in {\mathcal N} \}$.\\
	\hspace*{0.02in} {\bf Output:} Sets of RISs group: ${\mathcal K}_c$ for $c=1,\dots,C$.
	\label{alg:RISs-grouping}
	\begin{algorithmic}[1]

		\State Initialization:  ${\mathcal C}=\{1,\dots,C\}$, ${\mathcal K}_c=\emptyset$ for $c\in {\mathcal C}$, and ${\mathcal C}_{\rm tmp}=\emptyset$;

		\State{Calculate the average path loss from RIS$_k$ to the UEs in group $c
			\in{\mathcal C}$: $\bar \rho _{k,c}^{{\rm{group}}} = \sum\nolimits_{n \in {{\mathcal N}_c}} {\rho _{k,n}^{\rm{U}}/{N_c}} $;}
		\While{${\mathcal K}\ne \emptyset$}
		\State{Find the indices of the largest $\bar \rho _{k,c}^{{\rm{group}}}$: $\{ {{k^*},{c^*}} \}=\arg \mathop {\max }\limits_{k \in {\mathcal K},c \in {\mathcal C}\backslash {{\mathcal C}_{{\rm{tmp}}}}} \bar \rho _{k,c}^{{\rm{group}}}$}
		\State{Let ${{\mathcal K}_{{c^*}}} = {{\mathcal K}_{{c^*}}} \cup \{ {{k^*}} \}$, ${\mathcal K} = {\mathcal K}\backslash \{ {{k^*}} \}$;}
		\If{$| {{\mathcal K}_{{c^*}}} | = | {{\mathcal N}_{{c^*}}} |$}
		\State{${{\mathcal C}_{{\rm{tmp}}}}={{\mathcal C}_{{\rm{tmp}}}}\cup\{c^*\}$}
		\EndIf
		\EndWhile		
		\State  \Return  ${\mathcal K}_c$ for $c=1,\dots,C$.
	\end{algorithmic}
\end{algorithm}

\subsection{RIS-UE Association}\label{sec-4-3}
To provide each UE within a group with an independent data stream, it is necessary to select $K_c$ UEs from ${\mathcal N}_c$ for association with RISs in ${\mathcal K}_c$ as the rank of effective channel in group $c$ is limited by $K_c=|{\mathcal K}_c|$. The UE selection and RIS-UE association in each group impact UE interference. Since intra-group interference has been addressed through ZF precoding, we design the UE selection and RIS-UE association to minimize the maximal inter-group interference. This design can be formulated as
\begin{equation} \label{eq:R-U-asso}
	\begin{aligned}
		\mathop {\min }\limits_{\left\{ {	{\mathcal N}_c^{{\rm{sel}}}} \right\}_{c = 1}^C}& \mathop {\max }\limits_{n,m} ~{Q_{n,m}}\\
		{\rm s.t.}\qquad&{\mathcal N}_c^{{\rm{sel}}} \subseteq {	{\mathcal N}_c},~\left|{\mathcal N}_c^{{\rm{sel}}}\right|=K_c,~c = 1, \dots ,C, \\
		&n \in 	{\mathcal N}_c^{{\rm{sel}}},~m \in 	{\mathcal N}_{e}^{{\rm{sel}}},~c \ne e,
	\end{aligned}
\end{equation}
where ${\mathcal N}_c^{{\rm{sel}}}$ is the set of selected $K_c$ UEs in group $c$. Note that ${\mathcal N}_c^{{\rm{sel}}}$ is a sorted set, where the elements are associated one-to-one with the set ${\mathcal K}_c$ in order. A brute-force search approach is adopted to solve \eqref{eq:R-U-asso}.

\section{Numerical Results}\label{sec:5}

\begin{figure}
	\centering
	\includegraphics[width=0.5\textwidth]{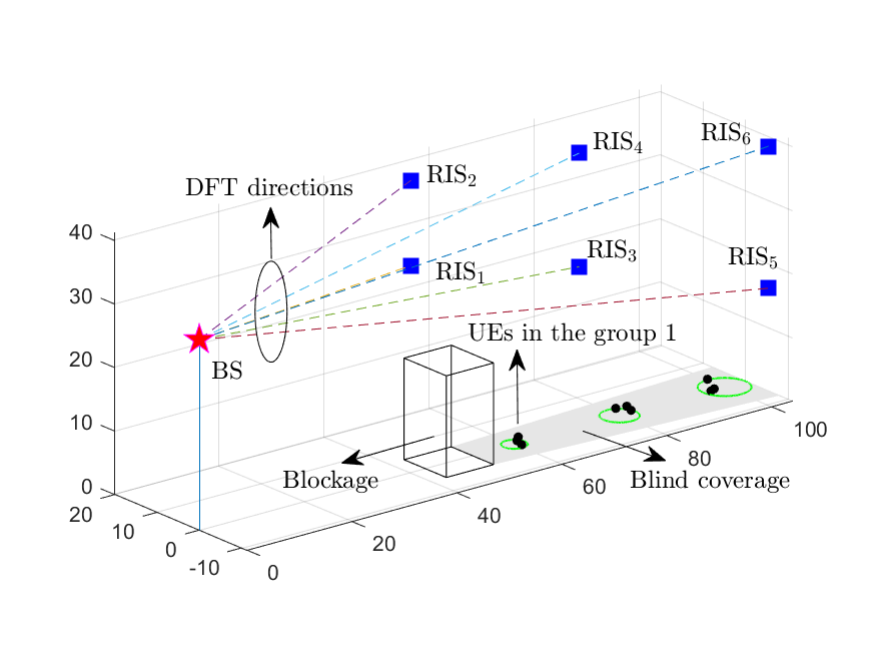}
	\caption{Overview of the multi-RIS assisted JSDM system.}
	\label{Fig.V-system}
	\vspace{-0.3cm} 
\end{figure}

In this section, we present numerical results to demonstrate the effectiveness of the proposed low-complexity channel BD and the performance of the RIS-assisted JSDM transmission. We provide visualizations of the effective channel and examine the impacts of the RIS size, number of data streams, deployment offset, and Rician factor on the SE performance. The simulations are conducted in an upper-6 GHz system with a carrier frequency of $f_c = 6.5$ GHz. Unless otherwise specified, the simulation system is configured with the following parameters: the BS is equipped with $M^{\rm B} = 9 \times 9$ antennas, and the RIS comprises $M^{\rm R} = 20 \times 20$ elements. As depicted in Fig. \ref{Fig.V-system}, the BS is positioned at $[0,0,30]$ of the Cartesian coordinates, where units are in meters. Due to blockage, the coverage hole is located on the $x$-axis. We consider $N = 9$ UEs within the coverage hole, assisted by $K = 6$ RISs. The UEs are grouped into $C = 3$ scattering rings, centered at $\{[60, 0, 0], [80, 0, 0], [100, 0, 0]\}$ with radii $\{2, 3, 4\}$, respectively. \textcolor{black}{Such UE distribution poses a significant challenge for JSDM without RISs' assistance, as it is difficult for the BS to distinguish UEs through horizontal spatial information. In this study, the RISs are introduced and aligned with the DFT directions of the BS.} The Rician factors, $\kappa^{\rm B}$ and $\kappa^{\rm U}$, are set to $10$ dB, reflecting dominant LoS propagation. The noise power is set at $-110$ dBm.

\begin{figure}
	\centering
	\includegraphics[width=0.5\textwidth]{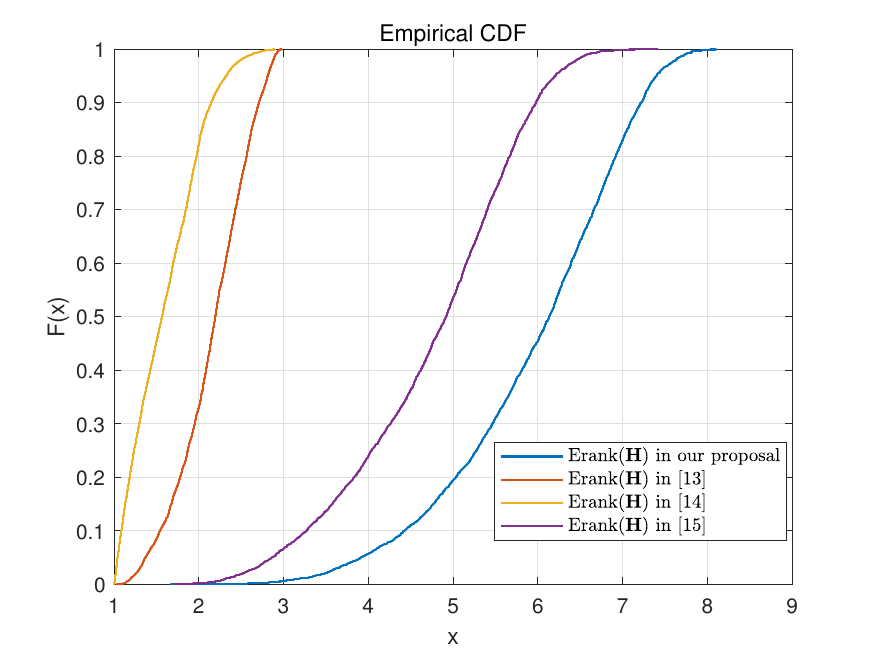}
	\caption{\textcolor{black}{CDF of the effective rank}.}
	\label{Fig.A-eRank}
\end{figure}

\textcolor{black}{In the considered scenario, we first compare the spatial multiplexing gain of the JSDM proposed in this study with those in \cite{JSDM-1RIS}--\cite{JSDM-TWC} by examining the effective rank of the multi-UE end-to-end channel. The effective rank of the channel ${\bf H}$ is defined as ${\rm {Erank}}({\bf H})={\rm {exp}}(-\sum\nolimits_{n} {\bar{\lambda}_n}{\rm {ln}}{\bar{\lambda}_n})$, where ${\bar \lambda _n} = \sqrt {{\lambda _n}} /\sum\nolimits_m {\sqrt {{\lambda _m}} } $ and ${{\lambda _n}}$ is the $n$-th singular value of ${\bf H}$ \cite{CC-mmwave}. For the single-RIS architecture considered in \cite{JSDM-1RIS}, six distributed RISs are consolidated into one at the position of RIS$_1$. For the two-hop RIS reflection model considered in \cite{JSDM-ICC}, the RIS$_5$ serves as the main RIS while the remaining RISs acting as intermediate RIS. For the architecture where one RIS serves one UE group, as considered in \cite{JSDM-TWC}, elements of RIS 2, 4, and 6 are merged into RIS 1, 3, and 5, respectively. The cumulative distribution function (CDF) of the effective rank under different JSDM architectures is presented in Fig. \ref{Fig.A-eRank}. In the RIS deployment schemes adopted in \cite{JSDM-1RIS} and \cite{JSDM-ICC}, the downlink signal is received by UEs after being reflected by a single RIS. As a result, the channels of different UEs are highly correlated, leading to a low effective rank of the multi-UE channel, which is almost always less than 3. The authors in \cite{JSDM-TWC} assigned a separate RIS to each UE group, thereby reducing the inter-group channel correlation. However, since the signals for UEs within the same group originate from the same RIS, the intra-group channel correlation remains high. Thus, the authors used different time resources to serves UEs within the same group. Unlike \cite{JSDM-1RIS}--\cite{JSDM-TWC}, we introduce multiple RISs to each UE group, fully leveraging the spatial degree of freedom of RISs to enhance the effective rank of the multi-UE channel and achieve greater spatial multiplexing gain.}

\begin{figure*}[t]
	\centering
	\includegraphics[width=0.8\textwidth]{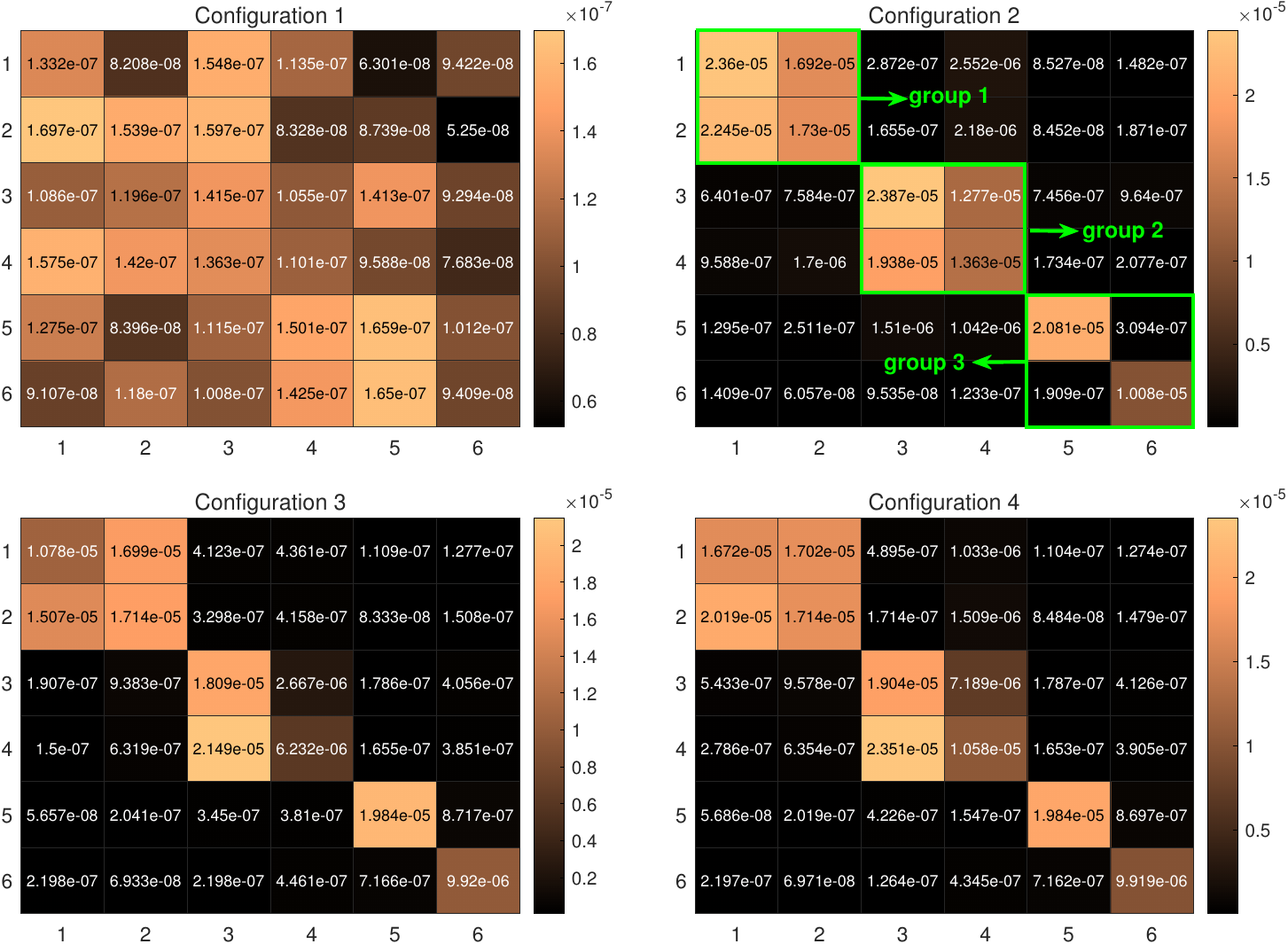}
	\caption{Visualization of the averaged effective channel matrix across different RIS configurations.}
	\label{Fig.A-400}
\end{figure*}
In the subsequent discussions, focusing on our proposal, the legends Configuration 1, 2, 3, and 4 refer to scenarios where the reflection coefficient vectors of RISs are designed using an identity matrix and by methods proposed in Sections \ref{sec:3.2.1}, B2, and B3, respectively. Under Configuration 1, RISs function passively, much like regular reflective surfaces, without actively manipulating signals to improve communication. Therefore, the system's performance under Configuration 1 is expected to be similar to that of a system without any RIS assistance, serving as a baseline for comparison with other configurations where the RIS actively enhances communication performance by customizing orthogonal group channels.

\subsection{Block Diagonalization}

Fig. \ref{Fig.A-400} visualizes the element amplitudes of the effective channel ${\bf H}^H{\bf F}$, averaged over 100 realizations. \textcolor{black}{As discussed in Section \ref{sec:3.1}, the rank of the effective channel is constrained by the number of RISs. For this reason, we select 6 UEs to form a $6 \times 6$ effective channel matrix.}  Each row pixel in the figure represents the amplitudes of the effective channel coefficients for each UE, with the global diagonal element indicating the target channel coefficient, and the off-diagonal elements representing intra-group and inter-group interference. \textcolor{black}{In Configuration 1,  amplitudes of the effective channel  are significantly low, approximately  $10^{-7}$, and the structure of the effective channel appears irregular. In contrast, in Configurations 2--4, where RISs are tailored to customize orthogonal group channels, the effective channel is reshaped with a well-block-diagonalized structure.}  \textcolor{black}{Notably, Configurations 3 and 4, which implement beam biasing to suppress the interference-to-target ratio, also result in a reduction of the target channel coefficients. This reduction leads to a degradation in the intra-group signal-to-interference ratio when compared to Configuration 2. For instance, the intra-group signal-to-interference ratio of UE$_4$, defined as ${\mathbb{E}}\{{[{\bf H}^H{\bf F}]}_{4,4}\}$/${\mathbb{E}}\{{[{\bf H}^H{\bf F}]}_{4,3}\}$, drops from 0.70 in Configuration 2 to 0.45 and 0.29 in Configurations 4 and 3, respectively.} \textcolor{black}{As shown in Fig. \ref{Fig.V-system}, the distribution radius of UEs increases from group 1 to group 3, indicating a decreasing channel correlation among intra-group UEs. This characteristic is embodied in the effective channel matrices of Configurations 2--4: the matrix for group 1 exhibits larger off-diagonal elements, while the matrix for group 3 becomes more diagonalized, reflecting a greater degree of orthogonality in the channel structure.}

\begin{figure}
	\centering
	\includegraphics[width=0.5\textwidth]{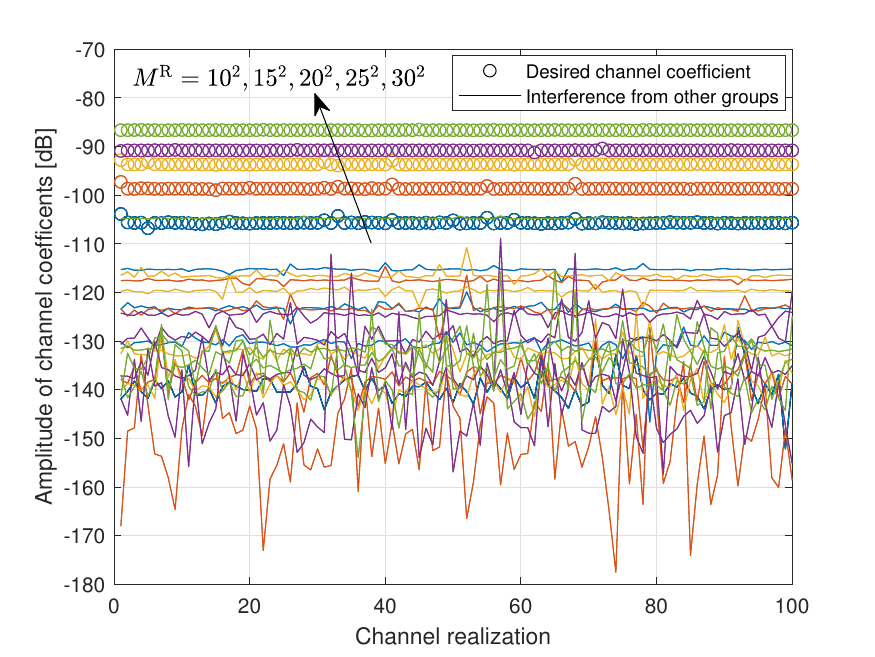}
	\caption{Amplitude of the desired and interference coefficients with different RIS sizes across 100 realizations.}
	\label{Fig.A-UE5}
	\vspace{-0.3cm}
\end{figure}

In the system under consideration, where 6 UEs are uniformly divided into 3 groups, each UE experiences interference from 4 UEs outside its group.  Fig. \ref{Fig.A-UE5} illustrates the amplitudes of the desired channel coefficient and the 4 inter-group interference coefficients for a UE under Configuration 3, with increasing RIS size across 100 realizations. \textcolor{black}{As the RIS array size increases, a notable enhancement in the amplitude of the desired channel coefficient is observed, driven by the higher array gain provided by the larger RIS. Additionally, the gap between the desired channel and interference coefficients widens, indicating that the increased number of RIS elements contributes to a more pronounced BD of the effective channel. This result is consistent with the analysis in Section \ref{sec:3.2.1} and suggests that with a larger RIS, the system can more effectively tailor the channel to minimize inter-group interference and improve overall signal quality.}

\subsection{Effect of the RIS Size}

\begin{figure}
	\centering
	\includegraphics[width=0.5\textwidth]{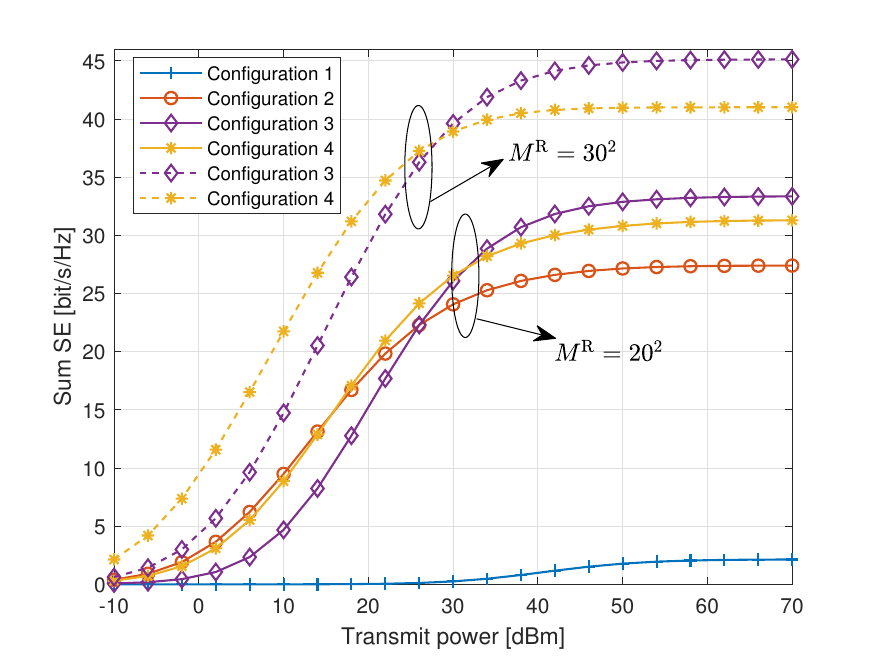}
	\caption{Sum SE versus transmit power with different RIS configurations for $N_c=2$.} \label{Fig.B-400}
	\vspace{-0.3cm}
\end{figure}
In this subsection, we evaluate the transmission performance of the proposed low-complexity JSDM in terms of SE. Fig. \ref{Fig.B-400} illustrates the trend of the sum SE as transmit power increases across different RIS configurations. \textcolor{black}{Given that 6 out of 9 UEs are selected and divided into 3 groups, each group contains $N_c=2$ UEs.} \textcolor{black}{The solid and dashed lines represent the cases with $M^{\rm R}$ being $20^2$ and $30^2$, respectively.} \textcolor{black}{Overall, there is a significant improvement in sum SE when RISs are configured, specifically in Configurations 2–4, to customize a block-diagonal effective channel for JSDM transmission.} The solid lines show that Configuration 3 achieves the highest sum SE at high transmit power levels, primarily because it is optimized to maximize the signal-to-interference ratio in interference-limited regimes. However, at lower transmit power levels, its performance lags behind Configurations 2 and 4. This is due to a decrease in the desired channel coefficient, as reflected in the diagonal element shown in Fig. \ref{Fig.A-400}. These results highlight the tradeoffs among the proposed configurations at different transmit power levels. \textcolor{black}{Given that Configurations 3 and 4 are designed for high and low transmit power levels, respectively, their performance will intersect at a specific transmit power value. To analyze how this intersection point varies with the RIS size, we present the sum SE of these two configurations with $M_{\rm R}$ increased from $20^2$ to $30^2$ in dashed lines. Notably, with an increased $M^{\rm R}$, the intersection point, shifts to lower transmit power levels.}  This shift is attributed to the asymptotic orthogonality effect, which causes the two configurations to exhibit similar performance in the high-dimensional regime, even at lower transmit powers.

\begin{figure}
	\centering
	\includegraphics[width=0.5\textwidth]{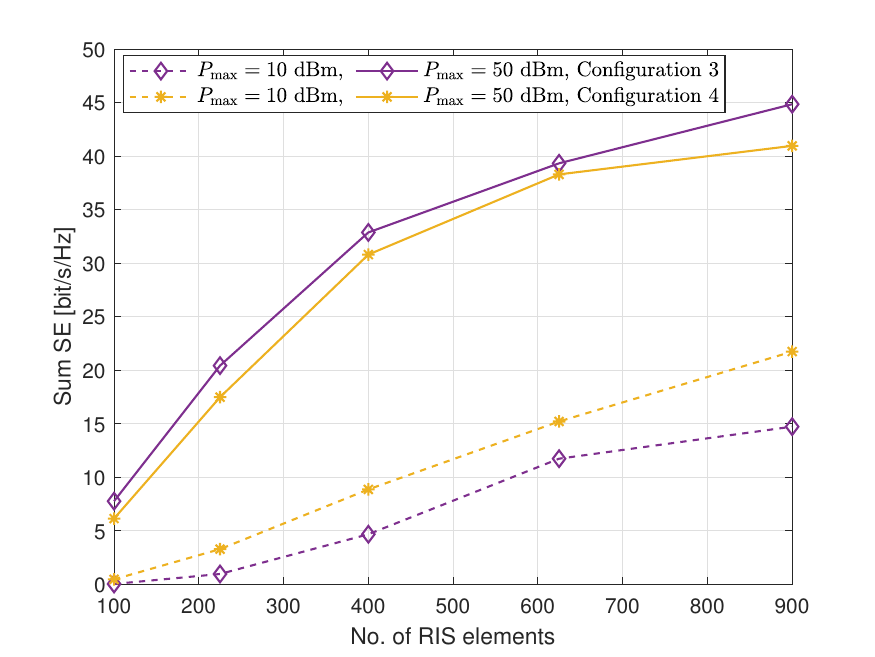}
	\caption{Sum SE comparison of Configurations 3 and 4 in noise- and interference-limited regimes with an increasing $M^{\rm R}$.}
	\label{Fig.B-MR}
	\vspace{-0.3cm}
\end{figure}

The performance of Configurations 3 and 4 is further analyzed in Fig. \ref{Fig.B-MR}. \textcolor{black}{As discussed in Section \ref{sec:3.2}, each configuration performs better in its respective regime: Configuration 3 excels in the interference-limited regime at $P_{\rm max} = 50$ dBm, while Configuration 4 is optimized for the noise-limited regime at $P_{\rm max} = 10$ dBm. In the interference-limited regime, as the number of RIS elements increases, Configuration 3 demonstrates a more pronounced improvement in sum SE, benefiting from the enhanced array gain that effectively mitigates interference. In contrast, in the noise-limited regime, the performance advantage shifts, with Configuration 4 outperforming Configuration 3. This occurs because the system becomes more sensitive to noise, and Configuration 4’s design is better suited to reduce noise-related impairments.}

\subsection{Effect of the Data Stream Number}

\begin{figure}
	\centering
	\includegraphics[width=0.5\textwidth]{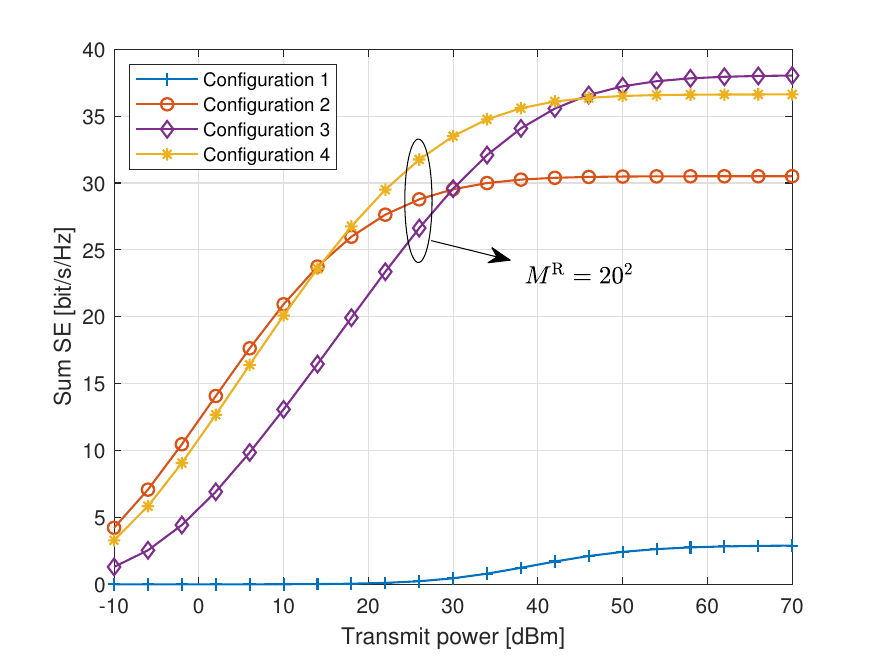}
	\caption{Sum SE versus the transmit power with different RIS configurations for $N_c=1$.} \label{Fig.C-F3}
\end{figure}

Unlike traditional JSDM, which requires careful optimization of the data stream number and PBF dimensions based on the dominant eigenmodes for each group, the proposed RIS-assisted JSDM simplifies implementation by setting both the number of data streams and the PBF dimension equal to the number of RISs. \textcolor{black}{When $K$ decreases from 6 to 3, we examine the impact of reducing the data stream number. As shown in Fig. \ref{Fig.C-F3}, Configurations 2--4 achieve higher sum SE compared to the solid lines in Fig. \ref{Fig.B-400}, despite the reduced number of RISs. This performance enhancement results from a reduction in interference, as fewer effective UEs are involved, leading to a cleaner channel with decreased intra-group and inter-group interference.} 
\begin{figure*}
	\centering
	\includegraphics[width=0.8\textwidth]{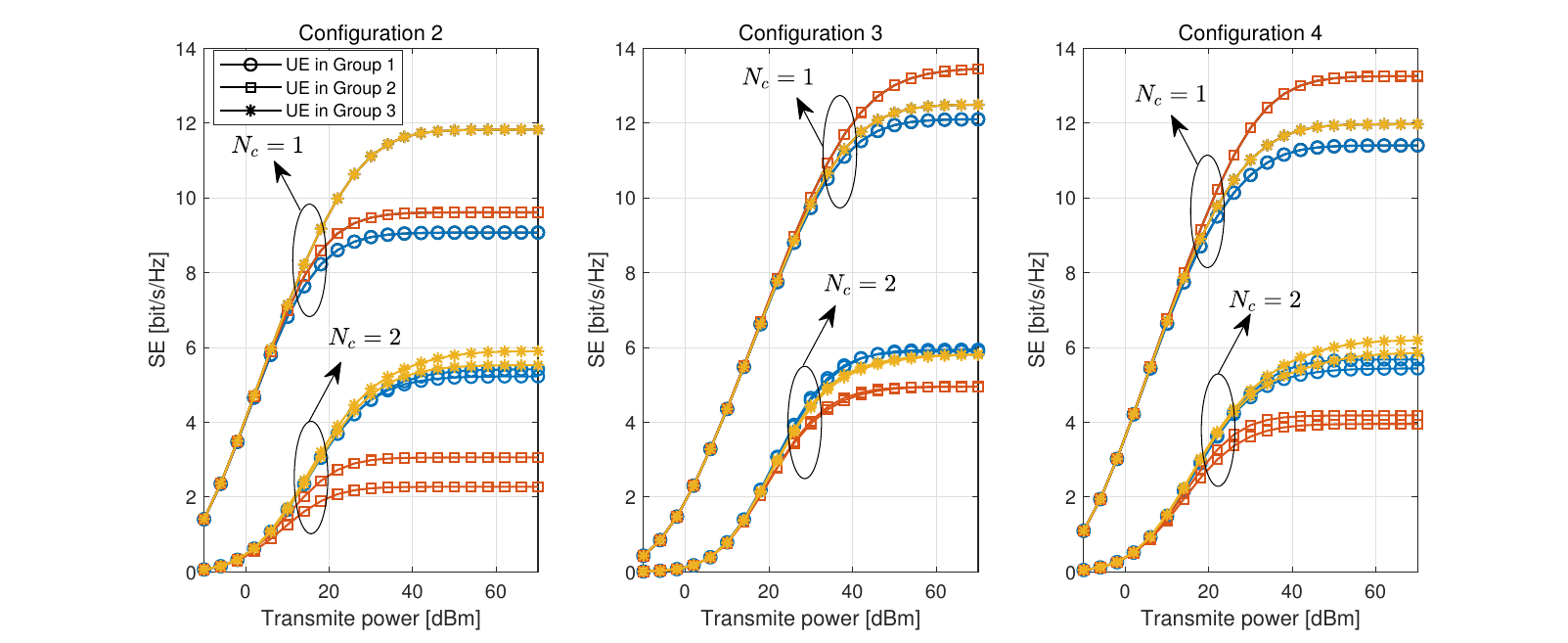}
	\caption{SE versus the transmit power for each UE with different RIS configurations.} \label{Fig.C-F3F6-single}
\end{figure*}
Fig. \ref{Fig.C-F3F6-single} presents a comparison of the SE for each UE under different RIS configurations. With a reduced number of effective UEs, the decrease in both intra-group and inter-group interference leads to substantial improvements in SE for each UE. \textcolor{black}{Notably, in Configuration 3, the observed reduction in the SE gap between UEs indicates that this configuration more effectively balances transmission quality across UEs, ensuring more equitable performance by minimizing maximal interference.}

\subsection{Effect of the Deployment \textcolor{black}{and Phase Shift} Offset}\label{sec:5.D}

\begin{figure}
	\centering
	\includegraphics[width=0.5\textwidth]{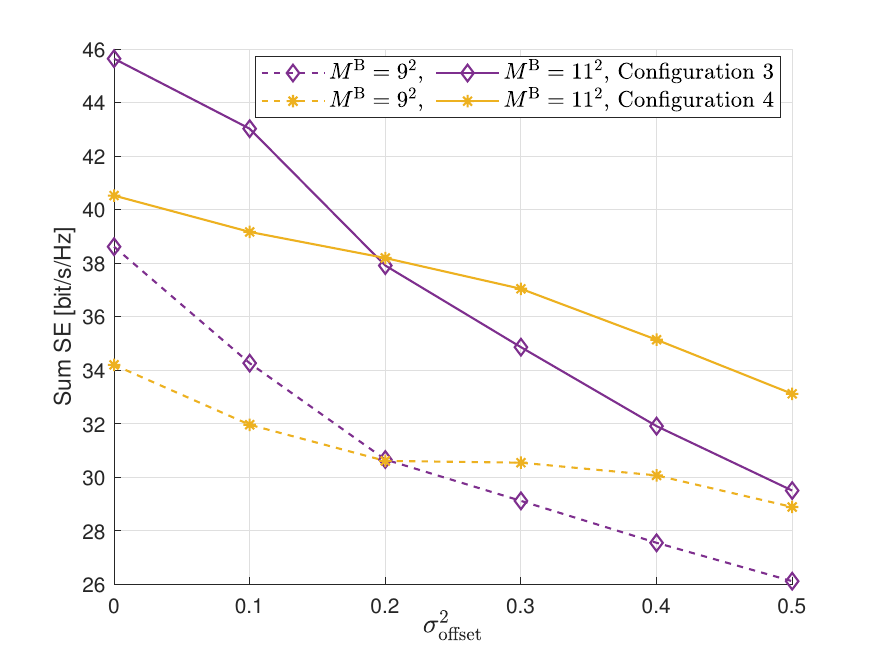}
	\caption{Sum SE versus the deployment offset of RISs with $P_{\rm max}=50$ dBm.}
	\label{Fig.D-9-11}
\end{figure}

Deploying the RIS at the DFT directions of the BS array is crucial for achieving orthogonal LoS transmission in the BS--RIS channel. \textcolor{black}{Any deviation in the RIS position disrupts this strict orthogonality, leading to additional interference between the data streams transmitted from the BS to the RISs.} In this section, the practical position of RIS$_k$ is modeled as ${\bf e}_{{\rm p},k}={\bf e}_{{\rm d},k}+{\bf e}_{{\rm o},k}$, where ${\bf e}_{{\rm d},k}$ represents the desired location and ${\bf e}_{{\rm o},k}\sim {\mathcal{ CN}}(0,{\sigma_{\rm offset}^2}{\bf I})$ models the offset. Fig. \ref{Fig.D-9-11} illustrates the impact of deployment offsets on the performance of Configurations 3 and 4, both of which are designed to reduce inter-group interference. As the offset increases, the performance of Configuration 3 deteriorates more rapidly than that of Configuration 4, due to more severe interference. Given that the DFT directions depend on the antenna array configuration of the BS, reducing the number of transmit antennas can increase the spacing between RISs, thereby potentially mitigating the performance degradation caused by deployment offset.

\textcolor{black}{Unlike deployment offsets that introduce inter-group interference into the BS-RIS channel, when phase shifts are offset, the RIS's reflected beam deviates from the intended direction to UEs, increasing inter-group interference in the RISs-UEs channel. Factors causing phase offsets mainly include the finite-bit quantization of RIS phase shifts and phase noise (PN) introduced by control circuits. When the RIS phase shift is subject to finite-bit quantization, the deviation between the actual phase shift of each element and the target phase shift is fixed, stemming from specific channel statistical information. In contrast, phase offsets due to PN, originating from the circuitry, are random errors. To evaluate the impacts of PN on the sum SE of the proposed JSDM in the interference-limited regime, Figure \ref{Fig.D-9-11} illustrates the performance of Configuration 3 versus varying PN variance at $P_{\rm max} = 50$ dBm. The practical reflection phase of the $n$-the element in the RIS$_k$ is modeled as ${\tilde\gamma}_{k,n}={\mathcal D}_b({\gamma}_{k,n})+n_{\rm PN}$, where ${\gamma}_{k,n}$  is the desired continuous phase shift (CPS), ${\mathcal D}_b(\cdot)$ represents the $b$-bits quantization, and $n_{\rm PN}\sim {\mathcal{CN}}(0,\sigma^2_{\rm PN}) $ is the PN with variance bing $\sigma^2_{\rm PN}$. Different solid lines represent different quantization bits, and the dashed lines represent cases without (w/o) PN. In discrete phase shift scenarios w/o PN, compared to CPS, the worst-case 1-bit quantization incurs a significant performance loss. However, as the number of quantization bits increases, the sum SE gap diminishes as expected. With the same level of  PN, the fewer the quantization bits, the less pronounced the sum SE degradation due to PN. This result is because the larger phase offsets resulting from fewer quantization bits partially counteract the phase offsets caused by PN.}
\begin{figure}
	\centering
	\includegraphics[width=0.5\textwidth]{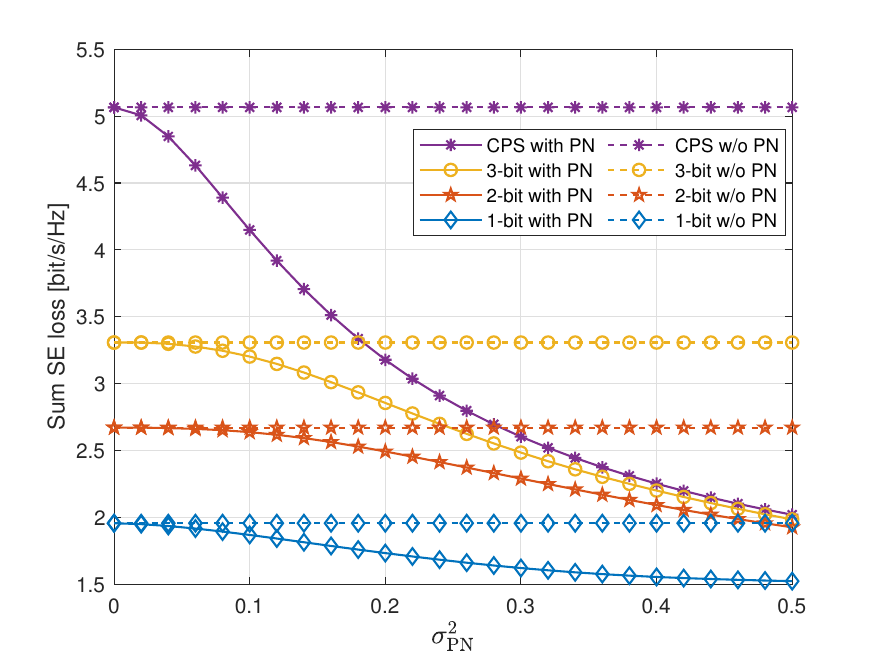}
	\caption{Sum SE versus the phase shift offset of RISs with $P_{\rm max}=50$ dBm. }
	\label{Fig.D1-1}
	\vspace{-0.3cm}
\end{figure}

\subsection{Effect of the Rician Factor}

\begin{figure}
	\centering
	\includegraphics[width=0.5\textwidth]{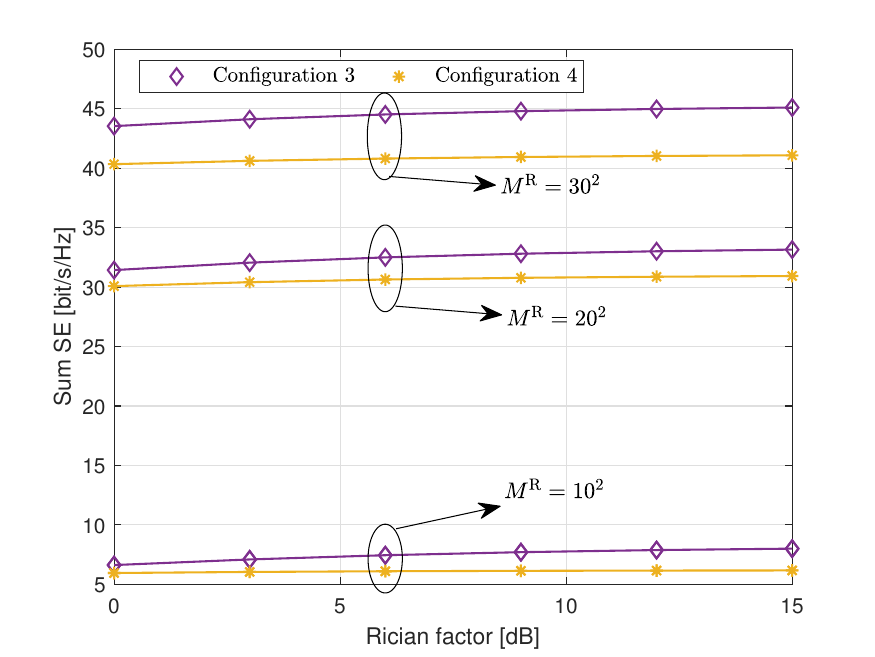}
	\caption{Sum SE versus the Rician factor of the BS--RISs channel.}
	\label{Fig.E-all}
	\vspace{-0.3cm}
\end{figure}

In Section \ref{sec:3.1}, we theoretically analyzed the rank of the end-to-end channel under the assumption that the link between the BS and RISs is dominated by LoS paths. \textcolor{black}{In other words, the Rician factor should be sufficiently large. Fig. \ref{Fig.E-all} shows that as the Rician factor increases from 0 dB to 15 dB, the sum SE remains relatively stable, demonstrating that the proposed RIS-assisted JSDM is robust across a wide range of Rician factors. As the Rician factor increases, Configuration 3 shows a growing advantage over Configuration 4, with the gap becoming more pronounced as the number of RIS elements increases. This suggests that a larger RIS further strengthens the dominance of the LoS path, improving the system's performance by mitigating the impact of NLoS components.}

\section{Conclusion}\label{sec:6}
In this study, we investigated JSDM in RIS-assisted FDD systems, where multiple UEs located in coverage holes are served via RIS-cascaded reflection links. Achieving BD of the dimension-reduced effective channel is crucial for JSDM implementation, as it minimizes inter-group interference. In conventional systems, where the end-to-end channel is uncontrollable, designing PBF for BD is complex and typically requires orthogonal eigen-matrices of the channel correlation matrices for different groups.
We introduced an innovative approach that leverages the adaptive nature of RIS to customize the end-to-end channel, thereby streamlining the BD process. Specifically, we investigated the strategic deployment of RIS positions and the configuration of their reflective properties to directly reduce inter-group interference in the BS–RIS and RIS–UE channels. Additionally, we developed an approximation of the channel cross-correlation coefficient to guide the association of RISs and UEs with the aim of reducing interference effectively.
Numerical results demonstrated that the effective channel can be well diagonalized through these methods. In terms of sum SE, the proposed low-complexity RIS-assisted JSDM was shown to achieve significant performance improvements in both noise-limited and interference-limited regimes.

\appendices
\section{}\label{App:A0}
\textcolor{black}{Utilizing the inner product result of ARVs in \cite[Appendix B]{ARV} for ${\bf{b}}_{k,0}^H{{\bf{b}}_{m,0}}$, we have 
\begin{equation}\label{eq-appA0-1}
	\begin{aligned}
		{\bf{b}}_{k,0}^H{{\bf{b}}_{m,0}} =\frac{{\sin \left( {{M^{{\rm{B}},{\rm{v}}}}{\Theta _{k,m}}} \right)}}{{{M^{{\rm{B}},{\rm{v}}}}\sin \left( {{\Theta _{k,m}}} \right)}}\frac{{\sin \left( {{M^{{\rm{B}},{\rm{h}}}}{\Phi _{k,m}}} \right)}}{{{M^{{\rm{B}},{\rm{h}}}}\sin \left( {{\Phi _{k,m}}} \right)}}
	\end{aligned},
\end{equation}
where ${\Theta _{k,m}} = \frac{\pi }{2}\cos ( {\theta _{m,0}^{{\rm{B}},{\rm{D}}}} ) - \frac{\pi }{2}\cos ( {\theta _{k,0}^{{\rm{B}},{\rm{D}}}} )$ and ${\Phi _{k,m}} = \frac{\pi }{2}\sin ( {\theta _{m,0}^{{\rm{B}},{\rm{D}}}} )\sin ( {\phi _{m,0}^{{\rm{B}},{\rm{D}}}} ) - \frac{\pi }{2}\sin ( {\theta _{k,0}^{{\rm{B}},{\rm{D}}}} )\sin ( {\phi _{k,0}^{{\rm{B}},{\rm{D}}}} )$. A straightforward implementation of ${\bf{b}}_{k,0}^H{{\bf{b}}_{m,0}} =0$ for $k\ne m$ is to install RISs at the different DFT directions of the BS's vertical antenna array \cite[Eqs. (27)--(28)]{CC-mmwave}. For instance, the vertical AoD of the LoS path in the BS--RIS$_k$ channel, ${\theta _{k,0}^{{\rm{B}},{\rm{D}}}}$, can be set to satisfy $\frac{\pi }{2}\cos ( {\theta _{k,0}^{{\rm{B}},{\rm{D}}}} )= \frac{\pi k}{{M^{\rm B,v}}}-\frac{\pi}{2}$. In this deployment, we have ${\Theta _{k,m}} = \frac{\pi (m-k)}{{M^{\rm B,v}}}$. When $k\ne m$,  $\sin( {{M^{{\rm{B}},{\rm{v}}}}{\Theta _{k,m}}} )=0$ always holds,  ensuring ${\bf{b}}_{k,0}^H{{\bf{b}}_{m,0}=0}$. }

\section{}\label{App:A}
\textcolor{black}{In the strong LoS dominant BS--RISs channel, where ${{\bf{B}}_k} = \beta _{k,0}^{\rm{B}}{{\bf{b}}_{k,0}}{\bf{r}}_{k,0}^H$}, the average of the element in the $n$-th row and $k$-column of the effective channel ${\bf H}^H{\bf F}$ can be expressed as
\begin{equation}\label{eq-APP-A1}
	{\mathbb{E}}\left\{ {\beta _{k,0}^{\rm{B}}{\left( {{\bf{u}}_{k,n}^H \odot {\bf{r}}_{k,0}^T} \right)}{\bm{\gamma}}_k^*} \right\} = {\beta}_{k,0}^{\rm{B}}{\mathbb{E}}\left\{ {{\left( {{\bf{u}}_{k,n}^H \odot {\bf{r}}_{k,0}^T} \right)}{\bm{\gamma}}_k^*} \right\}.
\end{equation}
Substituting \eqref{eq-Gamma} into \eqref{eq-APP-A1} and using the equation
\begin{equation}
	\begin{aligned}
		\left( {{{\bf{a}}^H} \odot {{\bf{b}}^T}} \right){\left( {{{\bf{c}}^*} \odot {\bf{d}}} \right)^*}= {{\bf{a}}^H}{\rm{diag}}\left( {{{\bf{d}}^*} \odot {\bf{b}}} \right){\bf{c}},
	\end{aligned}
\end{equation}
we can rewrite \eqref{eq-APP-A1} as
\begin{equation}\label{eq-APP-A2}
	 	{\mathbb{E}}\left\{ {\beta _{k,0}^{\rm{B}}\left( {{\bf{u}}_{k,n}^H \odot {\bf{r}}_{k,0}^T} \right){\bm{\gamma}}_k^*} \right\}
	 	 = \beta _{k,0}^{\rm{B}}{M^{\rm R}}{\mathbb{E}}\left\{ {{\bf{u}}_{k,n}^H{\bf{a}}\left( {{\Delta ^{{\rm{v}}}_k},{\Delta ^{{\rm{h}}}_k}} \right)} \right\},
\end{equation}
where ${\Delta ^{{\rm{v}}}_k} = \theta _{k,f( k )}^{\rm cntr} + \delta _{{\rm{v}},k}^{{\rm{bias}}}$,  ${\Delta ^{{\rm{h}}}_k} = \phi _{k,f( k )}^{\rm cntr} + \delta _{{\rm{h}},k}^{{\rm{bias}}}$, and $f(k)$ represents the group index of RIS$_k$. Expanding ${{\bf{u}}_{k,n}}=\sum\nolimits_{l = 0}^{{L_{k,n}}} {\beta _{k,n,l}^{\rm{U}}{\bf{r}}_{k,n,l}}$, the expectation in \eqref{eq-APP-A2} can be rewritten as
\begin{equation}\label{eq-APP-A3}
	\begin{aligned}
		&{\mathbb{E}}\left\{ {{\bf{u}}_{k,n}^H{\bf{a}}\left( {{\Delta ^{{\rm{v}}}_k},{\Delta ^{{\rm{h}}}_k}} \right)} \right\}
		= \sum\limits_{l = 0}^{{L_{k,n}}} {\mathbb{E}}{\left\{ {\beta _{k,n,l}^{{\rm{U,*}}}{\bf{r}}_{k,n,l}^H{\bf{a}}\left( {{\Delta ^{{\rm{v}}}_k},{\Delta ^{{\rm{h}}}_k}} \right)} \right\}} \\
		&= \sum\limits_{l = 0}^{{L_{k,n}}} {{\mathbb{E}}\left\{ {\beta _{k,n,l}^{{\rm{U,*}}}} \right\}{\mathbb{E}}\left\{ {{\bf{r}}_{k,n,l}^H{\bf{a}}\left( {{\Delta ^{{\rm{v}}}_k},{\Delta ^{{\rm{h}}}_k}} \right)} \right\}} .
	\end{aligned}
\end{equation}
Given that ${\mathbb{E}}{\{ {\beta _{k,n,l}^{{\rm{U,*}}}} \}}=0$ holds for $l>0$ and $\{{\beta _{k,n,0}^{{\rm{U}}}},{{\bf{r}}_{k,n,0}}\}$ are constant statistic CSI, \eqref{eq-APP-A3} can be simplified as
\begin{equation}
	{\mathbb{E}}\left\{ {{\bf{u}}_{k,n}^H{\bf{a}}\left( {{\Delta ^{{\rm{v}}}_k},{\Delta ^{{\rm{h}}}_k}} \right)} \right\} = \beta _{k,n,0}^{\rm{U}}{\bf{r}}_{k,n,0}^H{\bf{a}}\left( {{\Delta ^{{\rm{v}}}_k},{\Delta ^{{\rm{h}}}_k}} \right).
\end{equation}
Utilizing the inner product result of ARVs in Appendix B of \cite{ARV} for ${\bf{r}}_{k,n,0}^H{\bf{a}}( {{\Delta ^{{\rm{v}}}_k},{\Delta ^{{\rm{h}}}_k}} )$,  we finally obtain \eqref{eq-Pro-1}.

\section{}\label{App:B}

Given ${\bf h}_n={\sum\nolimits_{k = 1}^K {{{\bf{B}}_k}{{\bm{\Gamma}}_k}{{\bf{u}}_{k,n}}} }$, $Q_{n,m}$ can be unfolded as
\begin{equation}\label{eq-APP-B1}
	Q_{n,m} = \sum\limits_{k = 1}^K {\sum\limits_{t = 1}^K {\mathbb E}{\left\{ {{\bf{u}}_{t,m}^H{\bm{\Gamma}}_t^H{\bf{B}}_t^H{\bf{F}}{{\bf{F}}^H}{{\bf{B}}_k}{{\bm{\Gamma}}_k}{{\bf{u}}_{k,n}}} \right\}} }.
\end{equation}
By unfolding ${{\bf{u}}_{k,n}}=\sum\nolimits_{l = 0}^{{L_{k,n}}} {\beta _{k,n,l}^{\rm{U}}{\bf{r}}_{k,n,l}}$, the expectation in \eqref{eq-APP-B1} can be expressed as
\begin{equation}\label{eq-APP-B2}
	\begin{aligned}
		&{\mathbb E}\left\{ {{\bf{u}}_{t,m}^H{\bm{\Gamma}}_t^H{\bf{B}}_t^H{\bf{F}}{{\bf{F}}^H}{{\bf{B}}_k}{{\bm{\Gamma}}_k}{{\bf{u}}_{k,n}}} \right\}\\
		&= \sum\limits_{i = 0}^{{L_{t,m}}} {\sum\limits_{l = 0}^{{L_{k,n}}} {\mathbb E}{\left\{ {\beta _{t,m,i}^{{\rm{U,*}}}{\bf{r}}_{t,m,i}^H{\bm{\Gamma}}_t^H{\bf{B}}_t^H{\bf{F}}{{\bf{F}}^H}{{\bf{B}}_k}{{\bm{\Gamma}}_k}\beta _{k,n,l}^{\rm{U}}{{\bf{r}}_{k,n,l}}} \right\}} }.
	\end{aligned}
\end{equation}
Considering that $\beta _{t,m,i}^{{\rm{U}}}$ and $\beta _{k,n,l}^{{\rm{U}}}$ are independent variables for NLoS paths with zero mean, \eqref{eq-APP-B1} can be simplified as
\begin{equation}\label{eq-APP-B3}
{Q_{n,m}}= \sum\limits_{k = 1}^K {\beta _{k,m,0}^{\rm{U}}\beta _{k,n,0}^{\rm{U}}{\bf{r}}_{k,m,0}^H{\bm{\Gamma}}_k^H{\mathbb E}\left\{ {{\bf{B}}_k^H{\bf{F}}{{\bf{F}}^H}{{\bf{B}}_k}} \right\}{{\bm{\Gamma}}_k}{{\bf{r}}_{k,n,0}}} .
\end{equation}
Substituting ${{\bf{B}}_k}=\sum\nolimits_{l = 0}^{{L_k}} {\beta _{k,l}^{\rm{B}}{\bf{b}}_{k,l}{{\bf{r}}^H_{k,l}}}$ into \eqref{eq-APP-B3}, the expectation can be expanded as
\begin{equation}\label{eq-APP-B4}
	{\mathbb E}\left\{ {{\bf{B}}_k^H{\bf{F}}{{\bf{F}}^H}{{\bf{B}}_k}} \right\} = \sum\limits_{i = 0}^{{L_k}} {\sum\limits_{l = 0}^{{L_k}} {\mathbb E}{\left\{ {\beta _{k,i}^{{\rm{B,*}}}\beta _{k,l}^{\rm{B}}{{\bf{r}}_{k,i}}{\bf{b}}_{k,i}^H{\bf{F}}{{\bf{F}}^H}{{\bf{b}}_{k,l}}{\bf{r}}_{k,l}^H} \right\}} } .
\end{equation}
Because $\{\beta _{k,l}^{\rm{B}}\}_{l=1}^{L_{k}}$ are independent and identically distributed with zero-mean, we have
\begin{equation}\label{eq-APP-B5}
	\begin{aligned}
		&{\mathbb E}\left\{ {{\bf{B}}_k^H{\bf{F}}{{\bf{F}}^H}{{\bf{B}}_k}} \right\}=\sum\limits_{l = 0}^{{L_k}} {{\mathbb E}\left\{ {{{\left| {\beta _{k,l}^{\rm{B}}} \right|}^2}} \right\}{\mathbb E}\left\{ {{{\bf{r}}_{k,l}}{\bf{b}}_{k,l}^H{\bf{F}}{{\bf{F}}^H}{{\bf{b}}_{k,l}}{\bf{r}}_{k,l}^H} \right\}}\\
		&= {\left| {\beta _{k,0}^{\rm{B}}} \right|^2}{{\bf{r}}_{k,0}}{\bf{b}}_{k,0}^H{\bf{F}}{{\bf{F}}^H}{{\bf{b}}_{k,0}}{\bf{r}}_{k,0}^H\\
            &+ \sum\limits_{l = 1}^{{L_k}} {{\mathbb E}\left\{ {{{\left| {\beta _{k,l}^{\rm{B}}} \right|}^2}} \right\}{\mathbb E}\left\{ {{{\bf{r}}_{k,l}}{\bf{b}}_{k,l}^H{\bf{F}}{{\bf{F}}^H}{{\bf{b}}_{k,l}}{\bf{r}}_{k,l}^H} \right\}}   .
	\end{aligned}
\end{equation}
In the BS--RIS$_k$ channel, AoAs and AoDs of the NLoS path are assumed to be uniformly distributed. With this property, \eqref{eq-APP-B5} is equivalent to
\begin{equation}\label{eq-APP-B6}
		{\mathbb E}\left\{ {{\bf{B}}_k^H{\bf{F}}{{\bf{F}}^H}{{\bf{B}}_k}} \right\}
	= {\left| {\beta _{k,0}^{\rm{B}}} \right|^2}{{\bf{r}}_{k,0}}{\bf{b}}_{k,0}^H{\bf{F}}{{\bf{F}}^H}{{\bf{b}}_{k,0}}{\bf{r}}_{k,0}^H+ {L_k}{{\mathbb E}\left\{ {{{\left| {\beta _{k,l}^{\rm{B}}} \right|}^2}} \right\}{\mathbb E}\left\{ {{{\bf{r}}_{k,l}}{\bf{b}}_{k,l}^H{\bf{F}}{{\bf{F}}^H}{{\bf{b}}_{k,l}}{\bf{r}}_{k,l}^H} \right\}}   .
\end{equation}
In the considered scenario where the Rician factor of the BS--RISs channel is sufficiently larger, the ratio of ${L_k}{\mathbb{E}}\{ {{{| {\beta _{k,l}^{\rm{B}}} |}^2}} \}$ to ${| {\beta _{k,0}^{\rm{B}}} |^2}$ satisfies
\begin{equation}\label{eq-APP-B7}
	\frac{{{L_k}{\mathbb{E}}\left\{ {{{ | {\beta _{k,l}^{\rm{B}}} |}^2}} \right\}}}{{{{ | {\beta _{k,0}^{\rm{B}}}|}^2}}} = \frac{1}{{{\kappa ^{\rm{B}}}}}\to 0.
\end{equation}
Therefore, we approximate ${\mathbb E}\{ {{\bf{B}}_k^H{\bf{F}}{{\bf{F}}^H}{{\bf{B}}_k}} \}$ to the first term in \eqref{eq-APP-B6}. With the PBF designed as ${\bf f}_{i}={\bf b}_{i,0}$ and the deployment of RIS that ensures ${\bf b}_{k,0}^H{\bf b}_{i,0}=0$ for $k\ne i$,  the component ${\bf{b}}_{k,0}^H{\bf{F}}$ in \eqref{eq-APP-B6} is a vector with all $0$ except for the $k$-th element being $1$. Thus, the approximation of ${\mathbb E}\{ {{\bf{B}}_k^H{\bf{F}}{{\bf{F}}^H}{{\bf{B}}_k}} \}$ can be simplified by
\begin{equation}\label{eq-APP-B8}
	{\mathbb E}\left\{ {{\bf{B}}_k^H{\bf{F}}{{\bf{F}}^H}{{\bf{B}}_k}} \right\}\approx {\left| {\beta _{k,0}^{\rm{B}}} \right|^2}{{\bf{r}}_{k,0}}{\bf{r}}_{k,0}^H
\end{equation}
Substituting \eqref{eq-APP-B8} into \eqref{eq-APP-B3}, we finally obtain 
\textcolor{black}{
	\begin{equation}\label{eq-Qnm_app}
		{Q_{n,m}} \approx   \sum\nolimits_{k = 1}^K {{\beta _{k,n,m}}{\bf{r}}_{k,m,0}^H{\bm{\Gamma}}_k^H{{\bf{r}}_{k,0}}{\bf{r}}_{k,0}^H{{\bm{\Gamma}}_k}{{\bf{r}}_{k,n,0}}} ,
	\end{equation}
	where ${\beta _{k,n,m}} = {| {\beta _{k,0}^{\rm{B}}} |^2}\beta _{k,n,0}^{\rm{U}}\beta _{k,m,0}^{\rm{U}}$. It is evident that ${Q_{n,m}}$ is primarily determined by the LoS properties of UE$_n$ and UE$_m$, as well as the reflection vectors. When UE$_n$ and UE$_m$ belong to different groups, we have designed the reflection vectors for three cases in Section \ref{sec:3.2} to reduce channel cross-correlation from the perspective of effective channel BD. In the first case, BD is achieved through the asymptotic orthogonality of the ARVs. When the size of the RIS is insufficient, the approximate BD is the least effective, as the design of beam biases $(\delta _{{\rm{v}},k}^{{\rm{bias}}},\delta _{{\rm{h}},k}^{{\rm{bias}}})$ does not consider interference elimination. In this worst-case scenario, we analyze ${Q_{n,m}}$. We assume that RIS$_k$ is assigned to the UE$_k$, that is, ${{\bm{\gamma}}_k} = {M^{\rm R}}( {{\bf{r}}_{k,k,0}^*  \odot  {{\bf{r}}_{k,0}}  } )$. Based on ${{\bm{\gamma}}_k} $, \eqref{eq-Qnm_app} can be expressed as
	\begin{equation}
		{Q_{n,m}} \approx\sum\nolimits_{k = 1}^K {{\beta _{k,n,m}}\left( {{\bf{r}}_{k,m,0}^H{{\bf{r}}_{k,k,0}}} \right)\left( {{\bf{r}}_{k,k,0}^H{{\bf{r}}_{k,n,0}}} \right)}.
\end{equation}}


\begin{small}
	
\end{small}
%
%
%
%
%
%
%
%
%
%
%
%
%
%
%
%
%
%
%
%
%
%
%
%
%
%
%
%
%
%
%

\end{document}